\documentclass[11pt,a4paper]{article}
\usepackage{jstylemarg}
\usepackage[utf8]{inputenc}
\usepackage{bm}

\newcommand{\bi}{\begin{itemize}}
\newcommand{\ei}{\end{itemize}}
\newcommand{\bea}{\begin{align}}
\newcommand{\eea}{\end{align}}
\newcommand{\be}{\begin{equation}}
\newcommand{\ee}{\end{equation}}

\newcommand{\pl}{{\partial}}

\newcommand{\tcb}{\textcolor{blue}}

\makeatletter
\renewcommand*\env@matrix[1][\arraystretch]{%
  \edef\arraystretch{#1}%
  \hskip -\arraycolsep
  \let\@ifnextchar\new@ifnextchar
  \array{*\c@MaxMatrixCols c}}
\makeatother

\author{Charlotte SLEIGHT}

\author{\quad Massimo TARONNA\footnote{Postdoctoral Researcher of the Fund for Scientific Research-FNRS Belgium.}}

\affiliation{Universit\'e Libre de Bruxelles and International Solvay Institutes\\
\noindent ULB-Campus Plaine CP231, 1050 Brussels, Belgium}

\emailAdd{charlotte.sleight@gmail.com, massimo.taronna@ulb.ac.be}

\title{\centering
\huge{Spinning Witten Diagrams}}

\abstract{We develop a systematic framework to compute the conformal partial wave expansions (CPWEs) of tree-level four-point Witten diagrams with totally symmetric external fields of arbitrary mass and integer spin in AdS$_{d+1}$. Employing this framework, we determine the CPWE of a generic exchange Witten diagram with spinning exchanged field. As an intermediate step, we diagonalise the linear map between spinning three-point conformal structures and spinning cubic couplings in AdS. As a concrete application, we compute all exchange diagrams in the type A higher-spin gauge theory on AdS$_{d+1}$, which is conjectured to be dual to the free scalar $O\left(N\right)$ model. Given a CFT$_d$, our results provide the complete holographic reconstruction of all cubic couplings involving totally symmetric fields in the putative dual theory on AdS$_{d+1}$.}

\begin{document}

\maketitle

\section{Introduction}

Conformal field theories (CFT) are among the most well studied examples of quantum field theories (QFT), and are also among the few which admit a simple non-perturbative definition. This is owing to the fact that conformal invariance fixes all 2pt and 3pt correlation functions up to numerical coefficients and spectrum, usually referred to as CFT data. Associativity of the conformal operator algebra then allows to reconstruct, in principle, all higher point correlation functions at the non-perturbative level. This intrinsic simplicity triggered the pioneering works \cite{Ferrara:1972uq,Mack:1973mq,Ferrara:1973yt,Ferrara:1973vz,Polyakov:1974gs,Ferrara:1974nf} centred on the idea that symmetry and quantum mechanics alone should suffice to fix the dynamics of a QFT. This is known as the Bootstrap Program. This approach proved to be very successful in the 80's in the context of 2d CFTs \cite{Belavin:1984vu}, but remained dormant for CFTs in $d>2$ until very recently with the emergence of new analytic and numerical methods \cite{Dolan:2000ut,Dolan:2003hv,Rattazzi:2008pe,Rychkov:2009ij,Caracciolo:2009bx,Dolan:2011dv,Paulos:2014vya,Simmons-Duffin:2015qma}. These have led to striking new numerical results for 3d CFTs \cite{ElShowk:2012ht,Kos:2016ysd}, and have further triggered new analytic results for the conformal bootstrap in various limits \cite{Fitzpatrick:2012yx,Komargodski:2012ek,Alday:2015ota,Alday:2016njk,Gopakumar:2016wkt,Guerrieri:2016whh}.

CFTs also play a pivotal role in the holographic dualities, and are conjectured to be dual to gravitational theories living in a higher-dimensional anti-de Sitter (AdS) space \cite{Maldacena:1997re,Gubser:1998bc,Witten:1998qj}. From a bottom up perspective, AdS/CFT maps bulk and boundary consistency into each other, repackaging the various kinematic building blocks in terms of bulk or boundary degrees of freedom. To some extent, without imposing any additional constraint, this is a kinematic re-writing of the same physical object in two different bases. It was further shown in \cite{Heemskerk:2009pn,ElShowk:2011ag} that, in the large $N$ limit, standard Feynman diagram expansion in the bulk does repackage solutions to the bootstrap at leading order in $\frac1{N}$. In particular, this repackaging is in terms of Witten diagrams. From the bulk perspective, the latter play the role of the building blocks in terms of which the observables of the theory are expressed -- in direct analogy with S-matrix elements. In this holographic picture, the main physical consistency requirement is the emergence of bulk locality, see e.g. \cite{Gary:2009ae,Heemskerk:2009pn,Fitzpatrick:2012cg,Bekaert:2015tva,Skvortsov:2015lja,Maldacena:2015iua,Taronna:2016ats,Bekaert:2016ezc,Taronna:2016xrm,Sleight:2016hyl,Belin:2016dcu} for an incomplete list of works in this direction.   

Holography thus naturally provides a reformulation of the bootstrap problem in terms of different types of building blocks, which have a neat physical interpretation. The link between these two pictures is the main subject of the present work. In particular, we explicitly diagonalise the map between spinning three-point conformal structures and CPWE expansion on the boundary, and the local spinning bulk cubic couplings and Witten diagrams in the bulk. At the level of four-point functions this draws upon the link \cite{Costa:2014kfa,Sleight:2016hyl} between the shadow formalism \cite{Ferrara:1972xe,Ferrara:1972ay,Ferrara:1972uq,Ferrara:1973vz} and the split representation of AdS harmonic functions \cite{PhysRevD.10.589}. At the level of three-point functions, given a CFT$_d$ our results provide the complete holographic reconstruction of all cubic couplings involving totally symmetric fields in the putative dual theory on AdS$_{d+1}$. Previous works on the holographic reconstruction of bulk interactions from CFT correlation functions include: \cite{Petkou:2003zz,Bekaert:2014cea,Bekaert:2015tva,Skvortsov:2015pea,Sleight:2016dba,Sleight:2016hyl} in the context of higher-spin holography, and more recently \cite{grossrosen} in the context of the SYK model.

A key motivation behind this work is that such a bulk repackaging of CFT objects may give new insights into the bootstrap program, potentially providing novel methods to solve the crossing equations (see \cite{Gopakumar:2016wkt,Gopakumar:2016cpb,Dey:2016mcs,Aharony:2016dwx} for recent progress in this direction). Furthermore, this may also shed light on the quest for understanding quantum gravity and which CFTs admit a well-defined gravitational dual.

In the process of diagonalising the map between boundary OPE coefficients and bulk cubic couplings, we identify the corresponding bases of bulk and boundary 3pt and 4pt structures, which, in this sense, appear to be naturally selected by holography. This allows us to systematically study tree-level four-point exchange amplitudes involving totally symmetric fields of arbitrary mass and spin, and seamlessly derive their CPWE. Our formalism builds upon, and extends, the approach and results of the previous works \cite{Costa:2014kfa,Bekaert:2014cea,Bekaert:2015tva,Sleight:2016hyl}, which considered four-point Witten diagrams with only external scalars.\footnote{See also the very recent \cite{Chen:2017yia} which also employed this approach to compute the CPWE of four-point exchange Witten diagrams with external scalars, but using a different form for the cubic vertex. Since the latter vertex is equivalent to the ones used in \cite{Costa:2014kfa,Bekaert:2014cea} up to total derivatives, the result is the same up to contact terms.} As a concrete application, we determine all four-point exchange diagrams in the type-A higher-spin gauge theory on AdS$_{d+1}$, whose complete cubic couplings have been recently been established in metric-like form in \cite{Sleight:2016dba,Sleight:2016xqq}. This application of our results extends the works \cite{Bekaert:2014cea,Bekaert:2015tva} to include external gauge fields of arbitrary integer spin.

Let us also mention a parallel approach to the decomposition of Witten diagrams into conformal partial waves, which has recently been developed in \cite{Hijano:2015zsa}.\footnote{This has origins in the AdS$_3$/CFT$_2$ literature \cite{Hartman:2013mia,Fitzpatrick:2014vua,Asplund:2014coa,deBoer:2014sna,Hijano:2015rla,Fitzpatrick:2015zha,Alkalaev:2015wia,Hijano:2015qja}, on Virasoro and ${\cal W}_N$ conformal partial waves from the bulk.} This is underpinned by what is known as the ``geodesic Witten diagram'', the bulk object which computes a single conformal partial wave. The latter is essentially an exchange Witten diagram, but the crucial difference being that one integrates the cubic vertices over geodesics, as opposed to the full volume of AdS. The original paper \cite{Hijano:2015zsa} considered the case of external scalars, which has more recently been generalised to external fields with arbitrary integer spin: First to a single spinning external leg in \cite{Nishida:2016vds}, and very recently to each leg having arbitrary integer spin in \cite{Castro:2017hpx,Dyer:2017zef,Chen:2017yia}. It would be instructive to employ the geodesic Witten diagram approach developed in \cite{Hijano:2015zsa} to reproduce the explicit results obtained in \S \tcb{\ref{subsec::cpwesexch}} for the CPWEs of spinning exchange Witten diagrams. A prescription for the latter very recently appeared in \cite{Castro:2017hpx}, together with some results for spin-$0$ and spin-$1$ exchange diagrams.

The outline is as follows: Section \S $\tcb{\ref{spincpwes}}$ we review the CPWE in the standard setting of CFT, with a particular focus on the shadow formalism. In Section \S $\tcb{\ref{subsub::cpwesexchd}}$ we detail the parallel story in the bulk. In particular, how the harmonic function decomposition of four-point Witten diagrams provides the link with the shadow formalism via the split representation. In \S \tcb{\ref{subsec::spinning3pt}} we review the computation \cite{Sleight:2016dba} of generic spinning three point Witten diagrams, and present a convenient explicit diagonal form of the linear map between three-point conformal structures and local bulk cubic couplings. In \S \tcb{\ref{subsec::cpwesexch}} we apply the latter results to compute the CPWE of a generic spinning exchange Witten diagram in AdS$_{d+1}$, and furthermore in \S \tcb{\ref{subsec::hsgauge}} consider exchange diagrams in the concrete setting of the type A minimal higher-spin gauge theory. Various technical details are relegated to appendices \tcb{\ref{app::cna}}, \tcb{\ref{app:imprcur}}, \tcb{\ref{app::trofcur}} and \tcb{\ref{sec::seedint}}.

\section{Conformal Partial Waves}
\label{spincpwes}

\subsection{The Conformal Partial Wave Expansion}

The CPWE of correlation functions of primary operators in CFT is a decomposition into contributions from each conformal multiplet.\footnote{I.e. each conformal partial wave re-sums the
contribution of the primary operator + all of its descendants to the correlator, and is thus labelled by the dimension $\Delta$ and spin $s$ of the primary operator.} 
As a simple illustrative example, let us first consider the CPWEs of correlation functions involving scalar primary operators ${\cal O}_i$. For a four-point function expanded in the ${\sf s}$-channel,\footnote{We use sans-serif font to denote the expansion channels, to be distinguished from the spin, $s$.} this reads
\begin{equation}\label{scalarcpwe}
\langle {\cal O}_{1}\left(y_1\right){\cal O}_{2}\left(y_2\right){\cal O}_{3}\left(y_3\right){\cal O}_{4}\left(y_4\right) \rangle = \sum_{{\cal O}_{\Delta,s}} {\sf c}_{{\cal O}_1{\cal O}_2 {\cal O}_{\Delta,s}}{\sf c}^{{\cal O}_{\Delta,s}}{}_{{\cal O}_3{\cal O}_4}W_{\Delta,s}\left(y_i\right).
\end{equation}
The functions $W_{\Delta,s}$ are the conformal partial waves. These are purely kinematical objects, fixed completely by conformal symmetry and only depend on the representations of the primary operators ${\cal O}_{\Delta,s}$ and ${\cal O}_i$ under the conformal group.
Each conformal partial wave in the expansion \eqref{scalarcpwe} is weighted by the coefficients of the operator ${\cal O}_{\Delta,s}$ in the ${\cal O}_1 \times {\cal O}_2$ and ${\cal O}_3 \times {\cal O}_4$ OPEs. The CPWE thus effectively disentangles the dynamical information, which depends on the theory under consideration, from the universal information dictated by conformal symmetry.

Owing to these defining features, the CPWE expansion has turned out to be a powerful tool. This is highlighted, for instance, by its pivotal role in the successes (\cite{Rattazzi:2008pe,Rychkov:2009ij,Poland:2010wg,Poland:2011ey,ElShowk:2012ht}, to name a few) of the conformal bootstrap program \cite{Ferrara:1973yt,Polyakov:1974gs}. But in spite of this, explicit formulas for conformal partial waves are scarce. For the scalar case \eqref{scalarcpwe} closed form expressions are only available in even dimensions \cite{Dolan:2000ut,Dolan:2003hv}, while in other cases CPWs are inferred via indirect methods, such as: recursion relations \cite{Dolan:2011dv,Costa:2011dw,Echeverri:2015rwa,Iliesiu:2015qra,Echeverri:2016dun,Hogervorst:2016hal} and efficient series expansions \cite{Hogervorst:2013sma,Hogervorst:2013kva,Costa:2016xah}. 

In the following section we review another indirect approach, which is convenient for the CPWE of correlators involving operators with spin -- as well as their Witten diagram counterparts. This is underpinned by the shadow formalism of Ferrara, Gatto, Grillo, and Parisi \cite{Ferrara:1972xe,Ferrara:1972ay,Ferrara:1972uq,Ferrara:1973vz}, and leads to an expression for conformal blocks for operators in arbitrary Lorentz representations as an integral of three-point conformal structures. This approach was first considered by Hoffmann, Petkou and R\"uhl in \cite{Hoffmann:2000tr,Hoffmann:2000mx} for external scalar operators (see also \cite{Dolan:2011dv}), and the idea revisited and results generalised in \cite{SimmonsDuffin:2012uy,Costa:2014rya,Rejon-Barrera:2015bpa}.

\subsection{Spinning Conformal Partial Waves}

To a given primary operator ${\cal O}_{\Delta,s}$, can be associated a dual (or shadow) operator\footnote{$I_{\mu \nu}\left(y\right)$ is the inversion tensor \begin{equation}
    I_{\mu\nu}\left(y\right) = \delta_{\mu \nu} - \frac{2y_{\mu}y_{\nu}}{y^2} \: ; \qquad z_1 \cdot I\left(y\right) \cdot z_2 = z_1 \cdot z_2 - 2 \frac{z_1 \cdot y\, z_2 \cdot y }{y^2}.
\end{equation} The Thomas derivative \cite{10.2307/84634} (see also \cite{Dobrev:1975ru})
\begin{equation}
{\hat \partial}_{z^i} = \partial_{z^i} - \frac{1}{d-2+2 z \cdot \partial_z} z_i \partial^2_z,
\end{equation}
accounts for tracelessness, i.e. $z^2=0$.
}
\begin{equation}\label{shad}
{\tilde {\cal O}}_{\Delta,s}\left(y;z\right) = \kappa_{\Delta,s}\frac{1}{\pi^{d/2}}\int d^dy^\prime \frac{1}{\left(y-y^\prime\right)^{2d-2\Delta}} \left(z \cdot I(y-y^\prime) \cdot {\hat \partial}_{z^\prime}\right)^s  {\cal O}_{\Delta,s}\left(y^\prime;z^\prime\right),
\end{equation}
of the same spin and scaling dimension $d-\Delta$. The normalisation
\begin{equation}\label{kds}
    \kappa_{\Delta,s} =  \frac{\Gamma\left(d-\Delta+s\right)}{\Gamma\left(\Delta-\frac{d}{2}\right)} \frac{1}{\left(\Delta-1\right)_s},
\end{equation}
ensures that applying \eqref{shad} twice gives the identity. 

The key observation of the shadow approach to conformal partial waves is that the integral
\begin{align}
    {\cal P}_{\Delta,s} = \kappa_{d-\Delta,s} \frac{1}{\pi^{d/2}} \int d^dy\, {\cal O}_{\Delta,s} \left(y\right)|0\rangle  \langle 0 |{\tilde {\cal O}}_{\Delta,s} \left(y\right), \label{prsh}
\end{align}
projects onto the contribution of the conformal families of ${\cal O}_{\Delta,s}$ \emph{and its shadow} to a given four-point function. This is illustrated for the simplest case of scalar correlators in the following, before moving on to correlators of spinning operators.

\subsubsection{External scalar operators}
\label{subsubsec::extscalar}
Restricting, for now, to the  case of external scalar operators \eqref{scalarcpwe}, when projecting onto the {\sf s}-channel we have
\begin{align}
& \langle {\cal O}_{1}\left(y_1\right){\cal O}_{2}\left(y_2\right) {\cal P}_{\Delta,s}\,{\cal O}_{3}\left(y_3\right){\cal O}_{4}\left(y_4\right) \rangle\\  \nonumber
& \hspace*{4cm} = {\sf c}_{{\cal O}_1{\cal O}_2 {\cal O}_{\Delta,s}}{\sf c}^{{\cal O}_{\Delta,s}}{}_{{\cal O}_3{\cal O}_4} W_{\Delta,s}\left(y_i\right) + {\sf c}_{{\cal O}_1{\cal O}_2 {\tilde {\cal O}}_{\Delta,s}}{\sf c}^{{\tilde {\cal O}}}{}_{{\cal O}_3{\cal O}_4} W_{d-\Delta,s}\left(y_i\right),
\end{align}
which implies the following integral representation 
\begin{align}
& {\sf c}_{{\cal O}_1{\cal O}_2 {\cal O}_{\Delta,s}}{\sf c}^{{\cal O}_{\Delta,s}}{}_{{\cal O}_3{\cal O}_4} W_{\Delta,s}\left(y_i\right)\;+\; \text{shadow} \\ \nonumber
& \hspace*{2.75cm} = \kappa_{d-\Delta,s} \frac{1}{\pi^{d/2}} \int d^dy\, \langle {\cal O}_1\left(y_1\right){\cal O}_2\left(y_2\right){\cal O}_{\Delta,s}\left(y\right) \rangle \langle  {\tilde {\cal O}}_{\Delta,s}\left(y\right){\cal O}_3\left(y_3\right){\cal O}_4\left(y_4\right) \rangle,
\end{align}
for the total contribution as a product of two three-point functions. Stripping off the dynamical data leaves a universal integral expression for the sum of a conformal partial wave and its shadow, dictated purely by conformal symmetry and the operator representations:
\begin{align}\label{unishd}
& W_{\Delta,s}\left(y_i\right)\;+\;\text{shadow} \\ \nonumber
& \hspace*{2.2cm} =  \kappa_{d-\Delta,s} \frac{\gamma_{\tau,s}{\bar \gamma}_{\tau,s}}{\pi^{d/2}} \int d^dy\, \langle\langle {\cal O}_1\left(y_1\right){\cal O}_2\left(y_2\right) {\cal O}_{\Delta,s}\left(y\right) \rangle \rangle \langle\langle  {\tilde {\cal O}}_{\Delta,s}\left(y\right){\cal O}_3\left(y_3\right){\cal O}_4\left(y_4\right) \rangle \rangle,
\end{align}
where 
\begin{align}\label{gamma}
\gamma_{\tau,s} = \frac{\Gamma\left(\frac{d}{2}-\frac{\tau_3-\tau_4+\tau}{2}\right)}{\Gamma\left(\frac{\tau_3-\tau_4+\tau}{2}+s\right)}, 
\qquad {\bar \gamma}_{\tau,s} = \frac{\Gamma\left(\frac{d}{2}-\frac{\tau_4-\tau_3+\tau}{2}\right)}{\Gamma\left(\frac{\tau_4-\tau_3+\tau}{2}+s\right)}.
\end{align} 
The notation $\langle \langle \bullet \rangle \rangle $ denotes the kinematical part of the three-point function that is fixed by conformal symmetry. I.e. removal of the overall coefficient,\footnote{Using the definition \eqref{shad} one finds 
\begin{equation}
{\sf c}_{{\tilde {\cal O}}_{\Delta,s}{\cal O}_3{\cal O}_4} = \gamma_{\tau,s}{\bar \gamma}_{\tau,s}\,{\sf c}_{{\cal O}_{\Delta,s}{\cal O}_3{\cal O}_4}, 
\end{equation}
which is the origin of the factors \eqref{gamma} in the expression \eqref{unishd}.
}
\begin{subequations}\label{unit3pt}
\begin{align}
\langle {\cal O}_1\left(y_1\right){\cal O}_2\left(y_2\right){\cal O}_{\Delta,s}\left(y\right) \rangle & = {\sf c}_{{\cal O}_1{\cal O}_2 {\cal O}_{\Delta,s}}\langle\langle {\cal O}_1\left(y_1\right){\cal O}_2\left(y_2\right){\cal O}_{\Delta,s}\left(y\right) \rangle\rangle \\
\langle {\tilde {\cal O}}_{\Delta,s}\left(y\right){\cal O}_3\left(y_3\right){\cal O}_4\left(y_4\right) \rangle & = {\sf c}_{{\tilde {\cal O}}_{\Delta,s}{\cal O}_3{\cal O}_4} \langle\langle  {\tilde {\cal O}}_{\Delta,s}\left(y\right){\cal O}_3\left(y_3\right){\cal O}_4\left(y_4\right) \rangle \rangle,
\end{align}
\end{subequations}
which, for unit two-point function normalisation, is the removal of the OPE coefficients. Details on the above steps where given by Dolan and Osborn in \cite{Dolan:2011dv} section 3 and \cite{Dolan:2000ut}.

An integral expression for a single, non-shadow, conformal partial wave can be obtained by introducing a contour integral\footnote{The conformal partial wave $W_{\frac{d}{2}\pm i\nu,s}\left(y_i\right)$ decays exponentially for $\text{Im}\left(\nu\right) \rightarrow \mp \infty$. In applying the residue theorem to obtain the LHS from the RHS, for $W_{\frac{d}{2} \pm i\nu,s}$ we close the $\nu$-contour in the lower/upper half plane respectively.}
\begin{align} \label{contcpw}
 W_{\Delta,s}\left(y_i\right) & =  \frac{\left(\Delta-\frac{d}{2}\right)}{2\pi} \int^{\infty}_{-\infty}\frac{d\nu}{\nu^2 + \left(\Delta - \frac{d}{2}\right)^2} 
   \left(W_{\frac{d}{2}+i\nu,s}\left(y_i\right) +W_{\frac{d}{2}-i\nu,s}\left(y_i\right)\right),
 \end{align}
and inserting \eqref{unishd} into the integrand. The CPWE \eqref{scalarcpwe} can then be re-cast as a contour integral \cite{Dobrev:1975ru,1977LNP....63.....D}, 
\begin{align}\label{contcpwe}
&\langle {\cal O}_{1}\left(y_1\right){\cal O}_{2}\left(y_2\right){\cal O}_{3}\left(y_3\right){\cal O}_{4}\left(y_4\right) \rangle \\ \nonumber
& \hspace*{0.7cm} = \sum_{s} \int^{\infty}_{-\infty}d\nu\,  c_{s}\left(\nu\right) \int d^dy\, \langle\langle {\cal O}_1\left(y_1\right){\cal O}_2\left(y_2\right){\cal O}_{\tfrac{d}{2}+i\nu,s}\left(y\right) \rangle \rangle \langle\langle  {\cal O}_{\tfrac{d}{2}-i\nu,s}\left(y\right){\cal O}_3\left(y_3\right){\cal O}_4\left(y_4\right) \rangle \rangle,
\end{align}
where for ease of notation we defined ${\cal O}_{\frac{d}{2}-i\nu,s} = {\tilde {\cal O}}_{\frac{d}{2}+i\nu,s}$. 
The real function $c_{s}\left(\nu\right)$ encodes the dynamical information, with poles that carry the contribution from each spin-$s$ conformal multiplet. For example, a contribution from a conformal multiplet $\left[\Delta,s\right]$ manifests itself in $c_{s}\left(\nu\right)$ with a pole at $\tfrac{d}{2}+i\nu = \Delta$, with residue giving the OPE coefficients
\begin{equation}
 c_{s}\left(\nu\right) = \frac{\left(\Delta-\tfrac{d}{2}\right)\kappa_{d-\Delta,s}\gamma_{\tau,s}{\bar \gamma}_{\tau,s}}{2\pi^{d/2+1}} \frac{{\sf c}_{{\cal O}_1{\cal O}_2 {\cal O}_{\Delta,s}}{\sf c}^{{\cal O}_{\Delta,s}}{}_{{\cal O}_3{\cal O}_4}}{\left(\tfrac{d}{2}-\Delta+i\nu\right)\left(\tfrac{d}{2}-\Delta-i\nu\right)} + ...\,,
\end{equation}
where the $...$ denote possible contributions from other spin-$s$ multiplets in the spectrum. The contour integral form \eqref{contcpwe} of the CPWE admits a direct generalisation to four-point correlators involving operators with spin. The only difference with respect to the scalar case is that, in general, there is more than one conformal partial wave associated to each conformal multiplet. This is a consequence of the non-uniqueness of tensor structures compatible with conformal symmetry in three-point functions with more than one spinning operator. It is for this reason that external spinning operators are easily accommodated for in the integral form \eqref{contcpw} of the conformal partial wave, which we discuss in the following.

\subsubsection{Spinning Conformal Partial Waves}
\label{subsubsec::cpinningcpwe}
The integral representation \eqref{contcpw} of conformal partial waves carries over straightforwardly to CPWEs of four-point functions containing operators with spin. In this case, however, since the structure of three-point functions with more than one operator of non-zero spin is not unique, generally there is more than one conformal partial wave associated to the contribution of a given conformal multiplet.

The number of independent structures that may appear in a conformal three-point function with operators of spins $s_1$-$s_2$-$s_3$ is \cite{Metsaev:2005ar}
\begin{equation}\label{countmassive}
N\left(s_1,s_2,s_3\right) = \frac{\left(s_1+1\right)\left(s_1+2\right)\left(3s_2-s_1+3\right)}{6}-\frac{p\left(p+2\right)\left(2p+5\right)}{24}-\frac{1-\left(-1\right)^p}{16},
\end{equation}
where $s_1 \leq s_2 \leq s_3$ and $p \equiv \text{Max}\left(0,s_1+s_2-s_3\right)$. For correlation functions with two scalar operators there is just a single structure compatible with conformal symmetry, $N\left(0,0,s\right) = 1$, in accordance with the uniqueness of  conformal partial waves with external scalar operators that we previously observed.

Three-point functions involving two spinning operators have $N\left(s_1,s_2,s_3\right) > 1$. A general three-point function of spinning operators in a parity-even theory takes the form\footnote{\label{foo::sumn}To be more precise, $\sum\limits_{n_i} = \sum\limits^{\text{min}\left\{s_1,s_2\right\}}_{n_3=0}  \sum\limits^{\text{min}\left\{s_1-n_3,s_3\right\}}_{n_2=0}\sum\limits^{\text{min}\left\{s_2-n_3,s_3-n_2\right\}}_{n_1=0}$.\\

Let us also note that it is from this that one obtains the counting \eqref{countmassive}: $\sum\limits_{n_i} 1 = N\left(s_1,s_2,s_3\right)$.}
\begin{multline}\label{corr}
\langle {\cal O}_{\Delta_1,s_1}(y_1){\cal O}_{\Delta_2,s_2}(y_2){\cal O}_{\Delta_3,s_3}(y_3)\rangle\\=\sum_{n_i}{\sf c}^{n_1,n_2,n_3}_{s_1,s_2,s_3}\frac{{\sf Y}_1^{s_1-n_2-n_3}{\sf Y}_2^{s_2-n_3-n_1}{\sf Y}_3^{s_3-n_1-n_2}{\sf H}_1^{n_1}{\sf H}_2^{n_2}{\sf H}_3^{n_3}}{(y_{12}^2)^{\tfrac{\tau_1+\tau_2-\tau_3}{2}}(y_{23}^2)^{\tfrac{\tau_2+\tau_3-\tau_1}{2}}(y_{31}^2)^{\tfrac{\tau_3+\tau_1-\tau_2}{2}}}\,,
\end{multline}
with theory-dependent OPE coefficients ${\sf c}^{n_1,n_2,n_3}_{s_1,s_2,s_3}$. The six three-point conformally covariant building blocks are given by ($i \cong i+3$)
\begin{align}
 {\sf Y}_i & =\frac{z_i\cdot y_{i(i+1)}}{y_{i(i+1)}^2}-\frac{z_i\cdot y_{i(i+2)}}{y_{i(i+2)}^2}, \\
 {\sf H}_i &=\frac{1}{y_{(i+1)(i+2)}^2}\left(z_{i+1}\cdot z_{i+2}+\frac{2 z_{i+1}\cdot y_{(i+1)(i+2)}\,z_{i+2}\cdot y_{(i+2)(i+1)}}{y_{(i+1)(i+2)}^2}\right).
\end{align}
A conformal partial wave with spinning external operators is thus labelled by two three-component vectors ${\bf n} = \left(n_1,n_2,n\right)$ and ${\bf m} = \left(m,m_3,m_4\right)$, 
\begin{align}\label{mass123factor}
& W^{\text{{\bf n}},\text{{\bf m}}}_{\Delta,s}\left(y_i\right) \; + \; \text{shadow} \\ \nonumber
& \hspace*{.75cm} = {\kappa}_{d-\Delta,s}\frac{\gamma_{\tau,s} {\bar \gamma}_{\tau,s}}{\pi^{d/2}} \int d^dy\, \langle \langle {\cal O}_{\Delta_1,s_1}(y_1){\cal O}_{\Delta_2,s_2}(y_2) {\cal O}_{\Delta,s}(y)  \rangle \rangle^{(\text{{\bf n}})} \langle \langle {\tilde {\cal O}}_{\Delta,s}(y)  {\cal O}_{\Delta_3,s_3}(y_3){\cal O}_{\Delta_4,s_4}(y_4)\rangle \rangle^{(\text{{\bf m}})},
\end{align}
where, by applying the definition \eqref{unit3pt} of the operation $\langle\langle \bullet \rangle \rangle$, 
\begin{align} \label{strip123mass}
\langle \langle {\cal O}_{\Delta_1,s_1}(y_1){\cal O}_{\Delta_2,s_2}(y_2) {\cal O}_{\Delta_3,s_3}(y_3)  \rangle \rangle^{(\text{{\bf n}})}&=\frac{{\sf Y}_1^{s_1-n_2-n}{\sf Y}_2^{s_2-n-n_1}{\sf Y}_3^{s-n_1-n_2}{\sf H}_1^{n_1}{\sf H}_2^{n_2}{\sf H}_3^{n}}{(y_{12}^2)^{\tfrac{\tau_1+\tau_2-\tau}{2}}(y_{23}^2)^{\tfrac{\tau_2+\tau-\tau_1}{2}}(y_{31}^2)^{\tfrac{\tau+\tau_1-\tau_2}{2}}}.
\end{align}
 
In the same way, the shadow contribution can be projected out by introducing a contour integral as in \eqref{contcpw}.

\subsubsection{Spinning Conserved Conformal Partial Waves} \label{spinningwaves}
 
Conservation of external operators places additional constraints on conformal partial waves, which is a consequence of the conservation conditions on three-point functions of conserved operators \cite{Osborn:1993cr,Erdmenger:1996yc}. The latter relates the coefficients ${\sf c}^{n_1,n_2,n_3}_{s_1,s_2,s_3}$ in a general spinning three-point function \eqref{corr} amongst each other, reducing the number of independent forms to \cite{Metsaev:2005ar}
\begin{equation}
{\cal N}\left(s_1,s_2,s_3\right) = 1 + \text{min}\left\{s_1,s_2,s_3\right\},
\end{equation}
when each operator in the three-point function is conserved. The general form for a three-point function of conserved operators in $d>3$ is given as a generating functional by \cite{Stanev:2012nq,Zhiboedov:2012bm},\footnote{To extract the explicit structure of the correlator from the generating function form \eqref{gencc3pt} one expands and collects monomials of the form ${\sf Y}_1^{s_1-n_2-n_3}{\sf Y}_2^{s_2-n_1-n_3}{\sf Y}_3^{s-n_1-n_2}{\sf H}_1^{n_1}{\sf H}_2^{n_2}{\sf H}_3^{n_3}$.}
\begin{multline}
    \langle {\cal J}_{s_1}(y_1){\cal J}_{s_2}(y_2){\cal J}_{s_3}(y_3)\rangle  = \sum\limits^{\frac{1+\text{min}\left(s_1, s_2, s_3\right)}{2}}_{k=0}{\sf c}^{k}_{{\cal J}_{s_1}{\cal J}_{s_2}{\cal J}_{s_3}}\, {}_2F_{1}\left(\frac{1}{2}-k,-k,3-\frac{d}{2}-2k,-\frac{1}{2}\frac{{\sf \Lambda}}{{\sf H}^2_1{\sf H}^2_2{\sf H}^2_3}\right)  \\ \label{gencc3pt}
   \times\, \frac{e^{{\sf Y}_1 + {\sf Y}_2 + {\sf Y}_3} {}_0 F_{1}(d-2,-\frac{1}{2}{\sf H}_{1}){}_0 F_{1}(d-2,-\frac{1}{2}{\sf H}_{2}){}_0 F_{1}(d-2,-\frac{1}{2}{\sf H}_{3})}{(y_{12}^2)^{\tfrac{d}{2}-1}(y_{23}^2)^{\tfrac{d}{2}-1}(y_{31}^2)^{\tfrac{d}{2}-1}} {\sf \Lambda}^{2k},
\end{multline}
with
\begin{align}
    {\sf \Lambda} = {\sf Y}_1{\sf Y}_2{\sf Y}_3 + \frac{1}{2}\left[{\sf Y}_1{\sf H}_1+{\sf Y}_2{\sf H}_2 + {\sf Y}_3{\sf H}_3\right],
\end{align}
and $k$ takes both integer and half integer values. The OPE coefficients ${\sf c}^{k}_{{\cal J}_{s_1}{\cal J}_{s_2}{\cal J}_{s_3}}$ are not fixed by current conservation and depend on the theory. 

The above counting implies, for instance, that conserved conformal partial waves representing the contribution of a conserved primary operator are labelled by two half-integers $k \in \left\{0,1/2,1,...,1+\text{min}\left(s_1, s_2, s\right)/2\right\}$ and ${\tilde k} \in \left\{0,1/2,1,...,1+\text{min}\left(s,s_3, s_4\right)/2\right\}$, where $(s_1,s_2,s_3,s_4)$ are the spins of the external conserved operators,
\begin{multline}\label{consshad}
{\cal W}^{k,{\tilde k}}_{(s_1,s_2|s|s_3,s_4)}\left(y_i\right) \; + \; \text{shadow} \\
= {\kappa}_{d-\Delta,s}\frac{\gamma_{\tau,s} {\bar \gamma}_{\tau,s}}{\pi^{d/2}} \int d^dy\, \langle \langle {\cal J}_{s_1}(y_1){\cal J}_{s_2}(y_2){\cal J}_{s}(y)  \rangle \rangle^{(k)} \langle \langle {\tilde {\cal J}}_{s}(y){\cal J}_{s_3}(y_3){\cal J}_{s_4}(y_4) \rangle \rangle^{({\tilde k})},
\end{multline} 
 where ${\cal J}_{s}$ is the exchanged spin-$s$ conserved current and 
 \begin{multline}\label{consbasis}
 \langle \langle {\cal J}_{s_1}(y_1){\cal J}_{s_2}(y_2){\cal J}_{s}(y_3)  \rangle \rangle^{(k)} = {}_2F_{1}\left(\frac{1}{2}-k,-k,3-\frac{d}{2}-2k,-\frac{1}{2}\frac{{\sf \Lambda}}{{\sf H}^2_1{\sf H}^2_2{\sf H}^2_3}\right)  \\
 %\nonumber
 %  & \hspace*{3.25cm} 
   \times \frac{e^{{\sf Y}_1 + {\sf Y}_2 + {\sf Y}_3} {}_0 F_{1}(d-2,-\frac{1}{2}{\sf H}_{1}){}_0 F_{1}(d-2,-\frac{1}{2}{\sf H}_{2}){}_0 F_{1}(d-2,-\frac{1}{2}{\sf H}_{3})}{(y_{12}^2)^{\tfrac{d}{2}-1}(y_{23}^2)^{\tfrac{d}{2}-1}(y_{31}^2)^{\tfrac{d}{2}-1}} {\sf \Lambda}^{2k}.
 \end{multline}
 \\

Conservation of higher-spin currents is a powerful constraint, with the presence of a single exactly conserved current of spin $s > 2$ in the spectrum implying (in $d \ge 3$) that the theory is a free one \cite{Maldacena:2011jn,Boulanger:2013zza,Alba:2013yda,Alba:2015upa,Friedan:2015xea}.\footnote{Assuming a single stress tensor.} In this case the label $k$ of each independent structure in \eqref{gencc3pt} denotes the spin of the free conformal representation \cite{Zhiboedov:2012bm}. 
An example which we employ later on is the free scalar, where $k=0$ (the scalar singleton) and the corresponding conserved three-point structure can be conveniently expressed in terms of Bessel functions\footnote{To see this one employs the identity \begin{equation}
\Gamma\left(\alpha+1\right)x^{-\alpha}J_{\alpha}\left(2x\right) = 2^{-\alpha}{}_0F_{1}\left(\alpha+1;-\frac{x^2}{4}\right).
\end{equation}} 
\begin{align}\label{besslefon}
\langle {\cal J}_{s_1}(y_1){\cal J}_{s_2}(y_2){\cal J}_{s_3}(y_3)\rangle = {\sf c}^{0}_{{\cal J}_{s_1}{\cal J}_{s_2}{\cal J}_{s_3}}
\,\frac{\left(\prod_{i=1}^3\,2^{\tfrac{d}4-1}q_i^{\tfrac{1}{2}-\tfrac{d-2 }{4}}\Gamma(\tfrac{d-2}2)\,
J_{\frac{d}{2}-2}\left(\sqrt{q_i}\right) \right){\sf Y}_1^{s_1}{\sf Y}_2^{s_2}{\sf Y}_3^{s_3}}{(y_{12}^2)^{d/2-1}(y_{23}^2)^{d/2-1}(y_{31}^2)^{d/2-1}},
\end{align} 
where $q_i = 2 {\sf H}_i\,\partial_{{\sf Y}_{i+1}} \cdot \partial_{{\sf Y}_{i+2}}$. The OPE coefficients were worked out in \cite{Sleight:2016dba} to be 
 \begin{align}\label{onopecoef}
 {\sf c}^{0}_{{\cal J}_{s_1}{\cal J}_{s_2}{\cal J}_{s_3}} = N_{\text{d.o.f.}} {\sf c}^{0}_{s_1}{\sf c}^{0}_{s_2}{\sf c}^{0}_{s_3}, \qquad  \left({\sf c}^{0}_{s_i}\right)^2= \frac{\sqrt{\pi}\,2^{7-d-s_i}\,\Gamma(s_i+\tfrac{d-2}{2})\Gamma(s_i+d-3)}{N_{\text{d.o.f.}}\,s_i!\,\Gamma(s_i+\tfrac{d-3}{2})\Gamma(\tfrac{d-2}2)^2}\,.
 \end{align}

\section{CPWE of Spinning Witten Diagrams}
\label{subsub::cpwesexchd}

The integral representation of the CPWE is most suitable for establishing CPWEs of Witten diagrams, as it arises naturally from their harmonic function decomposition (see \cite{Sleight:2016hyl} for a detailed review):

The analogue of the CPWE expansion in the bulk is the decomposition into partial waves of the AdS isometry group. I.e. in terms of harmonic functions with energy and spin quantum numbers,\footnote{The harmonic function $\Omega_{\nu,s-2k}$ is a symmetric and traceless (in both sets of indices) spin $s-2k$ Eigenfunction of the Laplacian,
\begin{equation}
\left(\Box + \left(\tfrac{d}{2}+i\nu\right)\left(\tfrac{d}{2}-i\nu\right)+s-2k\right)\Omega_{\nu,s-2k} = 0,
\end{equation}
which is divergence-free, $\nabla \cdot \Omega_{\nu,s-2k}=0$.}
\begin{equation}
 \includegraphics[scale=0.45]{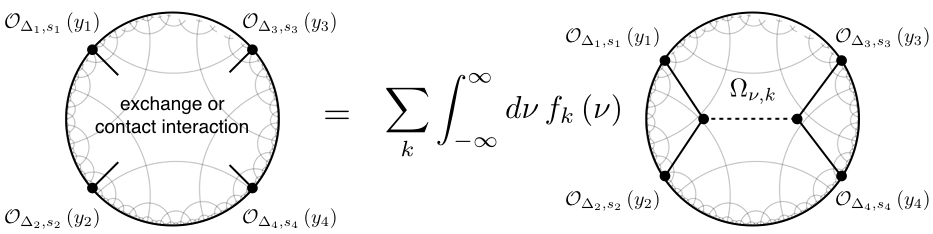},\label{harmfundecom}
\end{equation}

To make contact with the CPWE on the boundary, one notes that harmonic functions factorise \cite{PhysRevD.10.589}
\begin{equation} \label{splitharm}
    \Omega_{\nu,k}\left(x_1, u_1;x_2, u_2\right) = \frac{\nu^2}{\pi k!\left(\frac{d}{2}-1\right)_k} \int_{\partial\text{AdS}}d^dy\, \Pi_{\frac{d}{2}+i\nu,k}(x_1,u_1;y,{\hat \partial}_z) \Pi_{\frac{d}{2}-i\nu,k}\left(y,z;x_2,u_2\right),
\end{equation}
into a product of two boundary-to-bulk propagators of dimensions $\frac{d}{2}\pm i\nu$ and the same spin $k$. We see that, like for conformal partial waves (\S \tcb{\ref{spincpwes}} equation \eqref{contcpwe}), each bulk partial wave factorises into a product of two three-point Witten diagrams,
\begin{equation}\label{factor3ptwitt}
 \includegraphics[scale=0.45]{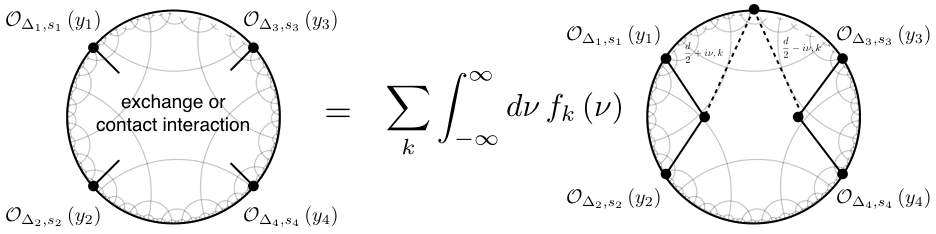}.
\end{equation}
Evaluating the bulk integrals yields a decomposition of the Witten diagram into products of three-point conformal structures on the boundary -- i.e. the integral representation \eqref{contcpwe} of the conformal partial wave expansion.

So far this approach has been applied to compute the CPWEs of tree-level Witten diagrams with only external scalars. This includes: The exchange of a massive spin-$s$ field and the graviton exchange \cite{Costa:2014kfa}\footnote{See also the very recent \cite{Chen:2017yia} which also employed this approach to compute the CPWE of four-point exchange Witten diagrams with external scalars, but using a different form for the cubic vertex. Since the latter vertex is equivalent to the ones used in \cite{Costa:2014kfa,Bekaert:2014cea} up to total derivatives, the result is the same up to contact terms.}; the exchange of spin-$s$ gauge field on AdS$_{d+1}$ \cite{Bekaert:2014cea} and contact diagrams for a general quartic scalar self-interaction \cite{Bekaert:2015tva}.

\subsection*{Spinning Exchange Witten Diagrams}

In this section we generalise the aforementioned results, to include all possible four-point exchange diagrams involving totally symmetric fields of arbitrary integer spin and mass -- both internally and externally.\footnote{For other works on spinning exchange diagrams, see: \cite{Francia:2007qt,Francia:2008hd} in the context of higher-spin gauge theories and more recently \cite{Castro:2017hpx} in the context of the geodesic Witten diagram decomposition of standard Witten diagrams.} To wit, we decompose into conformal partial waves the following general exchange of a spin-$s$ field of mass $m^2 R^2 =\Delta \left(\Delta-d\right)-s$ in AdS$_{d+1}$
\begin{equation}\label{exchgens1s2s3s4}
\includegraphics[scale=0.45]{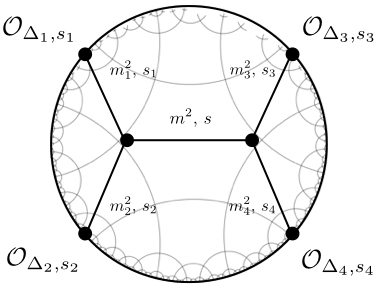},
\end{equation}
between external fields of spin $s_i$ and mass $m^2_i R^2 =\Delta_i \left(\Delta_i-d\right)-s_i$. 

The first step is to obtain the decomposition \eqref{harmfundecom} of the exchange diagram. This is achieved by expressing the bulk-to-bulk propagator of the exchanged field in a basis of harmonic functions \cite{Penedones:2007ns,Costa:2014kfa,Bekaert:2014cea}, which we review for massive fields in \S \tcb{\ref{subsec::massprop}} and for massless fields in \S \tcb{\ref{subssec::masslessprop}}. This leads to the decomposition \eqref{factor3ptwitt} of the exchange diagram \eqref{exchgens1s2s3s4} into products of tree-level three-point Witten diagrams, whose evaluation leads to the sought-for conformal partial wave expansion via identification with the integral form \eqref{mass123factor} of the conformal partial waves.

\subsection{Spinning three-point Witten diagrams}
\label{subsec::spinning3pt}
In the light of the decomposition \eqref{harmfundecom} of Witten diagrams, a key step to obtain CPWEs of spinning diagrams is therefore the evaluation of tree-level three-point Witten diagrams involving fields of arbitrary integer spin and mass. For parity even theories, this was carried out in \cite{Sleight:2016dba} in general dimensions, whose results we review here and also further supplement with new ones.

\subsubsection{Building blocks of cubic vertices}

Employing the ambient space formalism (reviewed in appendix \S \tcb{\ref{app::cna}}), a convenient basis of on-shell cubic vertices between totally symmetric fields $\varphi_{s_i}$ of spins $s_i$ and mass $m^2_i R^2 = \Delta_i\left(\Delta_i-d\right)-s_i$ is given by
\begin{align}\label{massivebasis}
& I^{n_1,n_2,n_3}_{s_1,s_2,s_3} = {\cal Y}^{s_1-n_2-n_3}_1{\cal Y}^{s_2-n_3-n_1}_2{\cal Y}^{s_3-n_1-n_2}_3 \\ \nonumber
& \hspace*{3cm} \times {\cal H}^{n_1}_1{\cal H}^{n_2}_2{\cal H}^{n_3}_3\, \varphi_{s_1}\left(X_1,U_1\right)\varphi_{s_1}\left(X_1,U_1\right)\varphi_{s_2}\left(X_2,U_2\right)\varphi_{s_s}\left(X_3,U_3\right)\Big|_{X_i=X},
\end{align}
which is parameterised by the six basic contractions
\begin{subequations}\label{6cont}
\begin{align}
    \mathcal{Y}_1&=\pl_{U_1}\cdot\pl_{X_2}\,,&\mathcal{Y}_2&=\pl_{U_2}\cdot\pl_{X_3}\,,&\mathcal{Y}_3&=\pl_{U_3}\cdot\pl_{X_1}\,,\\
    \mathcal{H}_1&=\pl_{U_2}\cdot\pl_{U_3}\,,&\mathcal{H}_2&=\pl_{U_3}\cdot\pl_{U_1}\,,&\mathcal{H}_3&=\pl_{U_1}\cdot\pl_{U_2}\,.
\end{align}
\end{subequations}
Recall that, in accordance with standard AdS/CFT lore, the basis elements \eqref{massivebasis} are in one-to-one correspondence with the independent three-point conformal structures \eqref{strip123mass}. 

The most general cubic vertex thus takes the form (c.f. footnote \ref{foo::sumn} for the sum over $n_i$)
\begin{equation}
V_{s_1,s_2,s_3} = \sum_{n_i}\, g^{n_1,n_2,n_3}_{s_1,s_2,s_3} I^{n_1,n_2,n_3}_{s_1,s_2,s_3}.
\end{equation}

The choice of basis \eqref{massivebasis} is convenient for three main reasons:

\begin{enumerate}
\item {\bf Simplicity:} The basis is built from the (commuting) ambient partial derivatives as opposed to the (non-commuting) AdS covariant derivatives.
\item {\bf Ease of manipulation and computation:} This is a consequence of the above simplicity. One important example is given by integration by parts in the ambient formalism.  While this is in general more involved compared to standard integration by parts directly on the AdS manifold, the basis \eqref{massivebasis} makes integration by parts as simple as in flat space. See \cite{Taronna:2012gb} for details on integration by parts in the ambient space framework, and in particular for the basis \eqref{massivebasis}.

\item {\bf Physical interpretation:} Any vertex expressed in terms of covariant derivatives can straightforwardly be cast in terms of the basis \eqref{massivebasis}, and vice versa, using (see \S \tcb{\ref{app::cna}})
\begin{equation}
\nabla_{A} = {\cal P}_{A}^{B}\frac{\partial}{\partial X^B}-\frac{X^B}{X^2}\Sigma_{AB},
\end{equation}
where 
\begin{equation}
\Sigma_{AB} = U_{\left[A\right.} \frac{\partial}{\partial U^{B\left.\right]}} =  U_{A} \frac{\partial}{\partial U^{B}} - U_{B} \frac{\partial}{\partial U^{A}},
\end{equation}
is the spin connection in the ambient generating function formalism. See appendix \tcb{B} of \cite{Sleight:2016dba} for more details about radial reduction.
\end{enumerate}

\subsubsection{Spinning Witten diagrams from a scalar seed}

Another virtue of the ambient space formalism is that Witten diagrams with spinning external legs can be seamlessly generated from those with only external scalars (which are comparably straightforward to evaluate) via the application of appropriate differential operators in the boundary variables. The ease of this approach to spinning Witten diagrams within the ambient framework is owing in particular to the homogeneity of the ambient representatives in both the bulk and boundary coordinates. The implication of this observation for the three-point Witten diagram generated by the basis vertex \eqref{massivebasis} is that it can be re-expressed in the form
\begin{equation}\label{iter}
\includegraphics[scale=0.425]{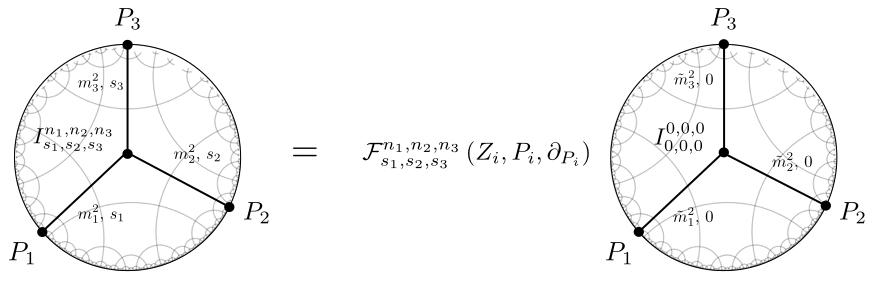},
\end{equation}
for some homogeneous differential operator ${\cal F}^{n_1,n_2,n_3}_{s_1,s_2,s_3}\left(Z_i,P_i,\partial_{P_i}\right)$, acting on the diagram generated by the coupling $I^{0,0,0}_{0,0,0}$ between scalars of some mass ${\tilde m}^2_i$. The latter is a well known integral which is straightforward to evaluate \cite{Freedman:1998tz}, which we review in \S \tcb{\ref{sec::seedint}}. Naturally, since the action of ${\cal F}^{n_1,n_2,n_3}_{s_1,s_2,s_3}$ increases the spin of the external legs, it will be a non-trivial function of $Z_i$. 

The decomposition \eqref{iter} of the spinning three-point Witten diagram can straightforwardly be obtained by noting that spinning bulk-to-boundary propagators have an analogous differential relationship to scalar bulk-to-boundary propagators \cite{Sleight:2016hyl}\footnote{By evaluating the action of the differential operator one recovers the standard expression \cite{Mikhailov:2002bp}
\begin{align}\label{buboprops}
    K_{\Delta,s}\left(X, U; P, Z \right) & = \left(U \cdot  {\cal P} \cdot Z \right)^s \frac{C_{\Delta,s}}{\left(-2 X \cdot P\right)^{\Delta}}, \qquad C_{\Delta,s} = \frac{\left(s+\Delta-1\right)\Gamma\left(\Delta\right)}{2 \pi^{d/2} \left(\Delta-1\right) \Gamma\left(\Delta+1-\tfrac{d}{2}\right)}.
    \end{align}
This also dictates the normalisation of the dual operator two-point function at large $N_{\text{d.o.f}}$
\begin{equation}\label{2ptnorm}
\langle {\cal O}_{\Delta,s}\left(y_1;z_1\right){\cal O}_{\Delta,s}\left(y_2;z_2\right) \rangle = \frac{C_{\Delta,s}}{\left(y^2_{12}\right)^{\Delta}}\left(z_1 \cdot z_2+\frac{2 z_1 \cdot y_{12} z_2 \cdot y_{21}}{y^2_{12}}\right)^s.
\end{equation}} 
\begin{equation}
K_{\Delta,s}\left(X,U;P,Z\right) = \frac{1}{\left(\Delta-1\right)_s}\left({\cal D}_P\left(Z;U\right)\right)^s K_{\Delta,0}\left(X;P\right),
\end{equation}
with differential operator
\begin{equation}
 {\cal D}_P\left(Z;U\right) = \left(Z \cdot U\right)\left(Z \cdot \frac{\partial}{\partial Z} - P \cdot \frac{\partial}{\partial P}\right) + \left(P \cdot U \right)\left(Z \cdot \frac{\partial}{\partial P}\right).
\end{equation}

Ambient partial derivatives of spinning bulk-to-boundary propagators, which arise naturally from the basis \eqref{massivebasis}, can readily be expressed in a similar form:
\begin{align}\label{ndiffbubulk}
\left(U_j \cdot \partial_X \right)^n K_{\Delta,s}\left(X,U_i;P,Z\right) & =  \frac{1}{\left(\Delta-1\right)_s}\left({\cal D}_P\left(Z;U\right)\right)^s \left(U_i \cdot \partial_X \right)^n K_{\Delta,0}\left(X;P\right)
\end{align}
with
\begin{equation}\label{partk}
\left( U_i \cdot \partial_{X} \right)^n K_{\Delta,0}\left(X; P \right)  = 2^n \left(\Delta\right)_n \left(U_i \cdot P\right)^n K_{\Delta+n,0}\left(X; P \right).
\end{equation}
This further illustrates the convenience of the choice of basis \eqref{massivebasis}.

Employing the expression for spinning bulk-to-boundary propagators \eqref{ndiffbubulk} one then obtains 

\begin{align}\label{fn1n2n3s1s2s3}
  & {\cal F}^{n_1,n_2,n_3}_{s_1,s_2,s_3} = \frac{2^{{\tilde s}_1+{\tilde s}_2+{\tilde s}_3}\left(\Delta_1\right)_{{\tilde s}_3}\left(\Delta_2\right)_{{\tilde s}_1}\left(\Delta_3\right)_{{\tilde s}_2}}{\left(\Delta_1-1\right)_{s_1}\left(\Delta_2-1\right)_{s_2}\left(\Delta_3-1\right)_{s_3}\left({\tilde s}_1\right)!\left({\tilde s}_2\right)!\left({\tilde s}_3\right)!}  \\ \nonumber
   & \hspace*{3cm} \times  {\cal H}^{n_1}_1{\cal H}^{n_2}_2{\cal H}^{n_3}_3  {\bar {\cal H}}^{{\tilde s}_2}_1 {\bar {\cal H}}^{{\tilde s}_3}_2 {\bar {\cal H}}^{{\tilde s}_1}_3 {\cal D}_{P_1}^{s_1}
  {\cal D}_{P_2}^{s_2}
   {\cal D}_{P_3}^{s_3}\left({\bar U}_1 \cdot P_1 \right)^{{\tilde s}_3}\left({\bar U}_2 \cdot P_2 \right)^{{\tilde s}_2}\left({\bar U}_3 \cdot P_3 \right)^{{\tilde s}_1}
\end{align}
where for concision we defined ${\tilde s}_i = s_i - n_{i-1}-n_{i+1}$ and introduced the auxiliary vector ${\bar U}_i$ which enters the contraction ${\bar {\cal H}}_i = \partial_{U_{i-1}} \cdot \partial_{{\bar U}_{i+1}}$. The mass of each scalar entering the seed vertex on the RHS of \eqref{iter} is given in terms of the quantum numbers of the original spinning fields on the LHS:
\begin{equation}
{\tilde m}^2_{i}R^2 = {\tilde \Delta}_i({\tilde \Delta}_i-d) \qquad \text{with} \qquad {\tilde \Delta}_i = \Delta_i+s_{i+2}-n_i-n_{i+1}.
\end{equation}

What remains to obtain the result for the spinning Witten diagram in the LHS of \eqref{iter} is to simply insert the result \eqref{123scal} for the scalar seed on the RHS and then act with the differential operator \eqref{fn1n2n3s1s2s3}. Denoting the amplitude by {\small $A^{n_1,n_2,n_3}_{s_1,s_2,s_3;\tau_1,\tau_2,\tau_3}$}, this procedure yields\footnote{The summation symbol is defined as:
\begin{equation}
\sum_{\alpha,\beta,\delta,\omega,\gamma}\equiv \sum_{\alpha_\kappa=0}^{s_\kappa-k_\kappa}\sum_{\beta_\kappa=0}^{k_{\kappa}}\sum_{\delta_{\kappa}=0}^{n_{\kappa}}\sum_{\omega_{\kappa}=0}^{\alpha_{\kappa-1}+\beta_{\kappa-1}}\sum_{\gamma_{\kappa}=0}^{\alpha_{\kappa-1}+\beta_{\kappa-1}}\,,
\end{equation}
} 

{\footnotesize
\begin{align}\label{6sums}
    &A^{n_1,n_2,n_3}_{s_1,s_2,s_3;\tau_1,\tau_2,\tau_3}\left(y_1,y_2,y_3\right) =   \text{P}_3 \sum_{\alpha,\beta,\delta,\omega,\gamma}\prod_{i=1}^3\, (-1)^{s_i-n_i-\delta_i +\alpha_i +\beta_i} 2^{s_i-n_i-\gamma_i  -\delta_i -\omega_i }\frac{n_i! (\alpha_i +\beta_i)! (s_i-n_{i+1} -n_{i-1} )! }{\gamma_i!\delta_i!\alpha_i!\omega_i!(\beta_i+\delta_{i+1} -n_{i+1}   +1)!}\nonumber\\\nonumber
    &\hspace{10pt} \times\frac{\left(\alpha_i +\beta_i+\Delta_i\right)_{s_i+\delta_{i(i+1)} -\gamma_{i+1} -n_{i+1}  -\omega_{i+1} -\Delta_i}}{(\alpha_i +\beta_i -\gamma_{i+1}  -\gamma_{i-1}  -\omega_{i+1}+1)! (s_i-\alpha_i -n_{i+1} -n_{i-1} -\omega_{i-1}  +1)!(n_{i+1} +n_{i-1}-\beta_i -\delta_{i+1}  -\delta_{i-1}  +1)!}\\\nonumber
    &\hspace{10pt} \times {\sf H}_1^{\gamma_1 +\delta_1 +\omega_1} {\sf H}_2^{\gamma_2 +\delta_2 +\omega_2 } {\sf H}_3^{\gamma_3 +\delta_3 +\omega_3 } {\sf Y}_1^{s_1-\gamma_2 -\gamma_3 -\delta_2 -\delta_3 -\omega_2 -\omega_3 } {\sf Y}_2^{s_2-\gamma_1 -\gamma_3 -\delta_1 -\delta_3 -\omega_1 -\omega_3 } {\sf Y}_3^{s_3-\gamma_1 -\gamma_2 -\delta_1 -\delta_2 -\omega_1 -\omega_2}\,,\nonumber
\end{align}
}
where $i \cong i+3$.

The pre-factor is given by
\begin{multline}
    \text{P}_3=\frac{1}{16\,\pi ^{d}}\frac1{(y_{12})^{\delta_{12}}(y_{23})^{\delta_{23}}(y_{31})^{\delta_{31}}}\,\Gamma \left(\sum_{\alpha}(\tfrac{\tau_\alpha}{2} +s_\alpha - n_\alpha ) -\tfrac{d}2\right)\prod_{i=1}^3\frac{ \Gamma ( \Delta_i-1)  ( \Delta_i+s_i -1) }{\, \Gamma \left(\Delta_i+1-\frac{d}{2}\right) }\,,\nonumber
\end{multline}
where
\begin{align}
    \delta_{(i-1)(i+1)}&=\frac12(\tau_{i-1}+\tau_{i+1}-\tau_i)\,,& \tau_i&=\Delta_i-s_i\,.
\end{align}

While the basis \eqref{massivebasis} is convenient as a means to evaluate spinning Witten diagrams, the resulting one-to-one map \eqref{6sums} between the bulk basis elements \eqref{massivebasis} and the canonical basis \eqref{strip123mass} of three-point conformal structures is rather involved. In the following section we introduce an alternative bulk and boundary pair of bases, through which the aforementioned bulk-to-boundary mapping simplifies dramatically and moreover allows to elegantly re-sum the expression \eqref{6sums}.

\subsubsection{A natural basis of cubic structures in AdS/CFT}

Let us motivate this alternative basis with a simple example. As observed in \cite{Sleight:2016dba}, the amplitude generated by the highest derivative basis vertex
\begin{equation}\label{hdvertex}
I^{0,0,0}_{s_1,s_2,s_3} = {\cal Y}^{s_1}_1{\cal Y}^{s_2}_2{\cal Y}^{s_3}_3 \, \varphi_{s_1}\left(X_1,U_1\right)\varphi_{s_1}\left(X_1,U_1\right)\varphi_{s_2}\left(X_2,U_2\right)\varphi_{s_s}\left(X_3,U_3\right)\Big|_{X_i=X},
\end{equation}
admits a very simple re-summation in terms of Bessel functions
\begin{multline} \label{1stbessl}
 A^{0,0,0}_{s_1,s_2,s_3;\tau_1,\tau_2,\tau_3}\left(y_1,y_2,y_3\right) =  \frac{B_{s_i;\tau_i}}{(y_{12})^{\delta_{12}}(y_{23})^{\delta_{23}}(y_{31})^{\delta_{31}}}\\\times \left[\prod_{i=1}^32^{\tfrac{\delta_{(i+1)(i-1)} }{2}-1} \Gamma \left(\frac{\delta_{(i+1)(i-1)} }{2}\right) q_i^{\frac{1}{2}-\frac{\delta_{(i+1)(i-1)}}{4}} J_{(\delta_{(i+1)(i-1)} -2)/2}\left(\sqrt{q_i}\right)\right]
\,{\sf Y}_1^{s_1}{\sf Y}_2^{s_2}{\sf Y}_3^{s_3},
\end{multline}
where we recall that {\small $q_i=2{\sf H}_i\pl_{{\sf Y}_{i+1}}\pl_{{\sf Y}_{i-1}}$} and the overall coefficient is given by
\begin{multline}
B_{s_i;\tau_i}=\frac{1}{16 \pi^d}\, \Gamma \left(\frac{\tau_1+\tau_2+\tau_3-d+2 (s_1+s_2+s_3)}{2} \right)\\\times\,\prod_{i=1}^3 \frac{\left(-2\right)^{s_i}\Gamma \left(s_i+\delta_{i(i+1)}\right)\Gamma \left(s_i+\delta_{(i-1)i}\right)\Gamma (s_i+\tau_i-1)}{\Gamma \left(s_i+\tau_i-\tfrac{d}{2}+1\right)\Gamma \left(\delta_{(i+1)(i-1)}\right)\Gamma (2 s_i+\tau_i-1)}.
\end{multline}
Such three-point conformal structures are for instance generated in free scalar CFTs (see e.g. \eqref{besslefon} for the case of three-point functions of conserved operators).

Given the simplicity and compactness of the three-point conformal structure \eqref{1stbessl} generated by the basis vertex \eqref{hdvertex}, it is temping to consider the following basis of conformal structures,

{\small \begin{align} \nonumber
&\left[\left[{\cal O}_{\Delta_1,s_1}(y_1){\cal O}_{\Delta_2,s_2}(y_2) {\cal O}_{\Delta_3,s_3}(y_3)\right]\right]^{(\text{{\bf n}})} \equiv\frac{{\sf H}_1^{n_1}{\sf H}_2^{n_2}{\sf H}_3^{n_3}}{(y_{12})^{\delta_{12}}(y_{23})^{\delta_{23}}(y_{31})^{\delta_{31}}} 
\left[\prod_{i=1}^32^{\tfrac{\delta_{(i+1)(i-1)} }{2}+n_i-1} \Gamma \left(\tfrac{\delta_{(i+1)(i-1)} }{2}+n_i\right)\right]
\\& \hspace*{2cm} \times\left[\prod_{i=1}^3 q_i^{\frac{1-n_i}{2}-\frac{\delta_{(i+1)(i-1)} }{4}} J_{(\delta_{(i+1)(i-1)}+2n_i -2)/2}\left(\sqrt{q_i}\right)\right]
\,{\sf Y}_1^{s_1-n_2-n_3}{\sf Y}_2^{s_2-n_3-n_1}{\sf Y}_3^{s_3-n_1-n_2} \label{nicebasis}
\end{align}}

\noindent
in the view of simplifying the map between bulk and boundary structures.

Indeed, working iteratively one finds that the conformal structure \eqref{nicebasis} is generated by the bulk vertex\footnote{For concision we define $\sum\limits_{m_i} = \sum\limits^{\text{min}\left\{s_1,s_2,n_3\right\}}_{m_3=0}  \sum\limits^{\text{min}\left\{s_1-n_3,s_3,n_2\right\}}_{n_2=0}\sum\limits^{\text{min}\left\{s_2-n_3,s_3-n_2,n_1\right\}}_{n_1=0}$}
\begin{equation}\label{Ical}
{\cal I}_{s_1,s_2,s_3}^{n_1,n_2,n_3}=\sum_{m_i} C_{s_1,s_2,s_3;m_1,m_2,m_3}^{n_1,n_2,n_3}\,I^{m_1,m_2,m_3}_{s_1,s_2,s_3}\,,
\end{equation}
with coefficients $C_{s_1,s_2,s_3;m_1,m_2,m_3}^{n_1,n_2,n_3}$ given by
\begin{multline}
   C_{s_1,s_2,s_3;m_1,m_2,m_3}^{n_1,n_2,n_3}= \left(\tfrac{d-2 (s_1+s_2+s_3-1)-\left(\tau_1+\tau_2+\tau_3\right)}{2} \right)_{m_1+m_2+m_3}\\\times\, \prod_{i=1}^3 \left[2^{m_i}\,\binom{n_i}{m_i} (n_i+\delta_{(i+1)(i-1)}-1)_{m_i}\right]\,,
\end{multline}
In particular, denoting the three-point amplitude generated by each basis element \eqref{Ical} by ${\cal A}^{n_1,n_2,n_3}_{s_1,s_2,s_3;\tau_1,\tau_2,\tau_3}$, we have:\footnote{Note, the vertices constructed here should not be confused with those written down in \cite{Dyer:2017zef,Chen:2017yia} in the context of geodesic Witten diagrams.}
\begin{equation}\label{IntegralBasisAdS}
\boxed{{\cal A}^{n_1,n_2,n_3}_{s_1,s_2,s_3;\tau_1,\tau_2,\tau_3}\left(y_1,y_2,y_3\right)={\sf B}(s_i;n_i;\tau_i)\,[[{\cal O}_{\Delta_1,s_1}(y_1){\cal O}_{\Delta_2,s_2}(y_2) {\cal O}_{\Delta_3,s_3}(y_3)]]^{(\text{{\bf n}})}}\,,
\end{equation}
with the coefficient ${\sf B}(s_i;n_i;\tau_i)$ given by
{\small \begin{multline}\label{beeeeeee}
{\sf B}(s_i;n_i;\tau_i)=\pi ^{-d}  (-2)^{(s_1+s_2+s_3)-(n_1+n_2+n_3)-4}\,\Gamma \left(\tfrac{\tau_1+\tau_2+\tau_3-d+2 (s_1+s_2+s_3)}{2} \right)\\  \times\,\prod_{i=1}^3 \frac{\Gamma \left(s_i- n_{i+1}+ n_{i-1}+\frac{\tau_i+\tau_{i+1}-\tau_{i-1}}{2}\right)\Gamma \left(s_i+n_{i+1}- n_{i-1}+\frac{\tau_i+\tau_{i-1}-\tau_{i+1}}{2}\right)\Gamma (s_i+n_{i+1}+n_{i-1}+\tau_i-1)}{\Gamma \left(s_i+\tau_i-\tfrac{d}{2}+1\right)\Gamma \left(2 n_i+\frac{\tau_{i+1}+\tau_{i-1}-\tau_i}{2}\right)\Gamma (2 s_i+\tau_i-1)}\,.
\end{multline}}

Given a CFT$_{d}$, the result \eqref{IntegralBasisAdS} provides the complete holographic reconstruction of all cubic couplings involving totally symmetric fields in the putative dual theory on AdS$_{d+1}$.\footnote{
 Previous works on the holographic reconstruction of bulk interactions from CFT correlation functions include: \cite{Petkou:2003zz,Bekaert:2014cea,Bekaert:2015tva,Skvortsov:2015pea,Sleight:2016dba,Sleight:2016hyl} for cubic and quartic couplings in the context of higher-spin holography. More recently cubic couplings have been extracted for the bulk dual of the SYK model in \cite{grossrosen}.}

\subsubsection*{Relation between bulk basis}
To conclude it is useful to spell out the explicit dictionary between the building blocks \eqref{massivebasis}, which allow to straightforwardly evaluate spinning Witten diagrams, and the basis \eqref{Ical} introduced in the previous section, which give a simple form for spinning three-point amplitudes.

Given a coupling of the form
\begin{equation}
\mathcal{V}_{s_1,s_2,s_3}=\sum_{n_i} g_{s_1,s_2,s_3}^{n_1,n_2,n_3}I_{s_1,s_2,s_3}^{n_1,n_2,n_3}\,,
\end{equation}
the problem is to determine the explicit form of the coefficient ${\tilde g}_{s_1,s_2,s_3}^{n_1,n_2,n_3}$ in the basis:
\begin{equation}
\mathcal{V}_{s_1,s_2,s_3}=\sum_{n_i}{\tilde g}_{s_1,s_2,s_3}^{n_1,n_2,n_3}{\cal I}_{s_1,s_2,s_3}^{n_1,n_2,n_3}\,,
\end{equation}
with $I_{s_1,s_2,s_3}^{n_1,n_2,n_3}$ and ${\cal I}_{s_1,s_2,s_3}^{n_1,n_2,n_3}$ given in \eqref{massivebasis} and \eqref{Ical}, respectively. Working iteratively, one arrives at the following expression for the coefficient ${\tilde g}_{s_1,s_2,s_3}^{n_1,n_2,n_3}$ as a function of the coefficients $g_{s_1,s_2,s_3}^{n_1,n_2,n_3}$ in the original basis:\footnote{For concision we define:
\begin{equation}
\sum_{m_i}=\sum_{m_3=n_3}^{\text{Min}\{s_1,s_2\}}\sum_{m_3
2=n_2}^{\text{Min}\{s_3,s_1-m_3\}}\sum_{m_1=n_1}^{\text{Min}\{s_2-m_3,s_3-m_2\}}
\end{equation}
}
\begin{equation}
{\tilde g}_{s_1,s_2,s_3}^{n_1,n_2,n_3}=\sum_{m_i}\Bigg[\tfrac{g_{m_1,m_2,m_3}}{ \left(\frac{d}{2}+1+\sum_{\alpha}(m_\alpha-s_\alpha-\tfrac{\tau_\alpha}2)\right)_{m_1+m_2+m_3}}\prod_{i=1}^3(-1)^{n_i+m_i}\frac{(2 n_i+\delta_{jk}-1)}{2^{m_i}\left( n_i+\delta_{jk}-1\right)_{m_i+1}}\binom{m_i}{n_i}\Bigg]\,,
\end{equation}
which is the inverse of the map \eqref{Ical}.

Notice that the new basis \eqref{nicebasis} generalises to non-conserved operators the basis \eqref{consbasis} of three-point conserved conformal structures. In this regard, our basis \eqref{nicebasis} seems to be naturally selected by free singleton CFTs.

\subsection{Spinning bulk-to-bulk propagators}

In this section we review previous works on the harmonic function decomposition of bulk-to-bulk propagators for totally symmetric fields of arbitrary mass and integer spin \cite{Bekaert:2014cea}.\footnote{For earlier works spinning bulk-to-bulk propagators, see \cite{Allen:1986dd,Allen:1985wd} by B. Allen for the graviton and (massive and massless) vector propagators (also \cite{Turyn:1988af,Liu:1998ty,DHoker:1998bqu,DHoker:1999bve,Costa:2014kfa}); for higher spin see \cite{Fronsdal:1978vb,Leonhardt:2003qu,Leonhardt:2003sn,Manvelyan:2005fp,Francia:2007qt,Francia:2008hd,Manvelyan:2008ks,Mkrtchyan:2010pp,Costa:2014kfa}.} \\

Up to cubic order in perturbations about the AdS background, a spin-$s$ field of mass {\small$m^2 R^2 = \Delta \left(\Delta-d\right)-s$} is governed by an effective Euclidean action of the form
\begin{equation}\label{genlag}
S_{m^2,s}\left[\varphi_s\right] = s! \int_{\text{AdS}} \frac{1}{2} \varphi_{s}\left(x,\partial_u\right)\left(\Box-m^2 + ...\right)\varphi_{s}\left(x,u\right) + \varphi_{s}\left(x,\partial_u\right) J_{s}\left(x,u\right) + {\cal O}\left(\varphi^4\right),
\end{equation}
where the source $J_s$ in the cubic interaction term is quadratic in the perturbations. The $...$ denote terms which depend on the off-shell completion, which we discuss case-by-case in the sequel.

Upon varying the action, the corresponding bulk-to-bulk propagator satisfies an equation of the form
\begin{equation}\label{buboeom}
\left(\Box_1-m^2 + ...\right)\Pi_{m^2,s}\left(x_1;x_2\right) = -\delta^{d+1}\left(x_1,x_2\right),
\end{equation}
where for convenience we suppressed the index structure, for now. To determine the propagator as a decomposition in harmonic functions, one can consider an ansatz of the form 
\begin{align} \label{ansatzbubu}
&\Pi_{m^2,s}\left(x_1,u_1;x_2,u_2\right) \\ \nonumber
& \hspace*{1.5cm} = \sum^{\lfloor s/2 \rfloor}_{k=0} \sum^{s-2k}_{l=0} \int^{\infty}_{-\infty} d\nu \, g_{k,l}\left(\nu\right) \left(u^2_1\right)^{k} \left(u^2_2\right)^{k} \left(u_1 \cdot \nabla_1\right)^{l} \left(u_2 \cdot \nabla_2\right)^{l} \Omega_{\nu,s-2k-l} \left(x_1,u_1;x_2,u_2\right).
\end{align}
The functions $g_{k,l}\left(\nu\right)$ are fixed by requiring that the equation of motion \eqref{buboeom} is satisfied.

We first review the solution for massive spinning fields before moving on to the massless case, where one has the additional requirement of gauge invariance.

\subsubsection{Massive case}
\label{subsec::massprop}

The Lagrangian formulation for freely propagating totally symmetric massive fields of arbitrary spin was first considered by Singh and Hagen in the 70's \cite{Singh:1974qz,Singh:1974rc}.\footnote{See also \cite{Rindani:1985pi,Rindani:1988gb,Aragone:1988yx,Zinoviev:2001dt}.} In order for the Fierz-Pauli physical state conditions \cite{10.2307/96758,Fierz:1939zz,Fierz:1939ix} 
\begin{align}\label{fp}
\left(\Box-m^2\right)\varphi_{s}\left(x,u\right)=0, \qquad
\left(\partial_u \cdot \nabla \right) \varphi_{s}\left(x,u\right)=0, \qquad
\left(\partial_u \cdot \partial_u \right) \varphi_{s}\left(x,u\right)=0,
\end{align}
to be recovered upon varying the action, the field content consists of the traceless field $\varphi_s$, and additional traceless auxiliary fields of ranks $s-2$, $s-3$, ..., $0$ which vanish on-shell.\footnote{See \cite{Francia:2007ee,Francia:2008ac} for an alternative formulation of the free massive Lagrangian in terms of curvatures, free from such auxiliary fields. Their removal, however, comes at the price of introducing non-localities.}

The complete off-shell form of the free Lagrangian is involved, and is moreover currently unavailable in its entirety on an AdS background. On the other hand, the terms which have not yet been identified explicitly are those which vanish on-shell (i.e. the ... in \eqref{genlag}) and thus only generate contact terms in exchange amplitudes. The latter are not universal contributions, as they are highly dependent on the field frame. For our purposes it is therefore not necessary to keep track of such terms,\footnote{When it is feasible we do keep track of contact terms, such as for the massless case introduced in the following section.} and we can solve the following equation for the massive spin-$s$ bulk-to-bulk propagator
\begin{equation}\label{trleemass}
\left(\Box_1-m^2\right)\Pi_{m^2,s}\left(x_1,u_1;x_2,u_2\right) = - \left\{ \left(u_1 \cdot u_2\right)^s\right\}\delta^{d+1}\left(x_1,x_2\right),
\end{equation}
where the notation $\left\{ \bullet \right\}$ signifies a traceless projection.

Since in this case the field is traceless, the following ansatz can be considered for the bulk-to-bulk propagator
\begin{equation}\label{massiveans}
\Pi_{m^2,s}\left(x_1,w_1;x_2,w_2\right) = \sum^{s}_{l=0} \int^{\infty}_{-\infty} d\nu \, g_{l}\left(\nu\right) \left(w_1 \cdot \nabla_1\right)^{l} \left(w_2 \cdot \nabla_2\right)^{l} \Omega_{\nu,s-l} \left(x_1,w_1;x_2,w_2\right),
\end{equation}
where the null auxiliary vectors $w^2_i=0$ enforce tracelessness. Substituting into the equation of motion \eqref{trleemass}, one finds \cite{Bekaert:2014cea}
\begin{align}
g_{l}\left(\nu\right) & = \frac{1}{\left(\frac{d}{2}-\Delta\right)^2+\nu^2-l+l(d+2s-\ell-1)} \\  \nonumber
& \hspace*{5cm} \times \frac{2^l \left(s-l+1\right)_{l}\left(\frac{d}{2}+s-l-\frac{1}{2}\right)_{l}}{l! \left(d+2s-2l-1\right)_{l}\left(\frac{d}{2}+s-l+i\nu\right)_l\left(\frac{d}{2}+s-l-i\nu\right)_l}.
\end{align}
Before moving on to consider the massless case, let us briefly highlight some generic features of the propagator \eqref{massiveans}:

\begin{itemize}
\item The traceless and transverse part of the propagator (corresponding to $l=0$ in \eqref{massiveans})
\begin{equation}\label{massivett}
\Pi^{TT}_{m^2,s}\left(x_1;x_2\right) = \int^{\infty}_{-\infty} \frac{d\nu}{\left(\frac{d}{2}-\Delta\right)^2+\nu^2} \Omega_{\nu,s}\left(x_1;x_2\right),
\end{equation}
is universal, and encodes the propagating degrees of freedom. 

\item The remaining contributions from harmonic functions of spin $< s$ (the $l > 0$ in \eqref{massiveans}) are purely off-shell, and generate only contact terms in exchange amplitudes. 
\end{itemize}

\subsubsection{Massless case}
\label{subssec::masslessprop}

On contrast to the massive case discussed in the previous section, the construction of free Lagrangians for massless fields is somewhat simplified owing to the additional guidance provided by gauge invariance. 

Recalling that the concept of masslessness in AdS is slightly deformed owing to the background curvature, requiring gauge invariance of the Fierz-Pauli system \eqref{fp} under the gauge transformation
\begin{equation}\label{spinsgt}
\delta_{\xi} \varphi_s\left(x,u\right) = \left(u\cdot \nabla \right) \xi_{s-1}\left(x,u\right),
\end{equation}
fixes $\Delta = s+d-2$ in the mass {\small $m^2 R^2 = \Delta \left(\Delta-d\right)-s$}. The complete off-shell Lagrangian form was determined by Fronsdal in the 70's \cite{Fronsdal:1978rb}, and reads
\begin{align}  \label{fronsdal0}
    S^{(2)}_{\text{Fronsdal}}\left[\varphi_s\right] & = \frac{s!}{2}  \int_{\text{AdS}_{d+1}}  \varphi_{s}\left(x;\partial_u\right) {\cal G}_s\left(x;u\right),
\end{align}
where ${\cal G}_s$ is the corresponding spin-$s$ generalisation of the linearised Einstein tensor
\begin{align}
   {\cal G}_s\left(x;u\right) & = \left(1-\frac{1}{4} \,u^2\, \partial_u \cdot \partial_u \right) \mathcal{F}_{s}\left(x; u, \nabla, \partial_u \right) \varphi_s\left(x, u\right),
\end{align}
with $\mathcal{F}_{s}$ the so-called Fronsdal operator 
\begin{align} \label{Fronsdaltensor}
{\cal F}_{s}(x,u,\nabla,\partial_u)
& =
\Box- m^2-u^2(\partial_u\cdot \partial_u)
-\;(u\cdot \nabla)\left((\nabla\cdot\partial_u)-\frac{1}{2}(u\cdot \nabla)
(\partial_u\cdot \partial_u)
\right).
\end{align}
The latter is fixed by invariance under linearised spin-$s$ gauge transformations \eqref{spinsgt} with symmetric and traceless rank $s-1$ gauge parameter $\xi_{s-1}$.\footnote{Alternative formulations have been developed which eliminate this algebraic trace constraint on the gauge parameter, however they come at the price of introducing introducing non-localities \cite{Francia:2002aa} or auxiliary fields \cite{Francia:2005bu,Francia:2007qt,Francia:2010ap}.} The Bianchi identity 
\begin{equation}\label{bianchi}
\left(\partial_u \cdot \nabla \right){\cal G}_s\left(x,u\right) = 0
\end{equation}
requires that the field $\varphi_s$ is double-traceless\footnote{Note that the double-trace of $\varphi_s$ is gauge invariant owing to the tracelessness of the gauge parameter. Forgoing the double-traceless constraint \eqref{dtconst} without introducing auxiliary fields (apart from deforming the Bianchi identity \eqref{bianchi} and thus requiring a modification of the action \eqref{fronsdal0}) would lead to the propagation of non-unitary modes, which one may try to kill by imposing appropriate boundary conditions. This has been shown to be possible in flat space \cite{Francia:2010ap} though it is not yet clear if this approach can be extended to AdS space-times, or if it is compatible with introducing a source. For this reason we stick to the standard Fronsdal formulation \eqref{fronsdal0} with double-trace constraint \eqref{dtconst}.}
\begin{equation}\label{dtconst}
\left(\partial_u \cdot \partial_u \right)^2 \varphi_{s}\left(x,u\right) = 0.
\end{equation}
To determine the bulk-to-bulk propagator one needs to invert the equation of motion with source
\begin{equation}\label{cueqm}
 (1-\frac{1}{4} \,u^2\, \partial_u \cdot \partial_u) \mathcal{F}_{s}\left(x; u, \nabla, \partial_u \right) \varphi_s\left(x, u\right) = - J_{s}\left(x,u\right),
\end{equation}
where from gauge-invariance it follows that $J_s$ is conserved on-shell, $\left(\partial_u \cdot \nabla \right)J_s\approx 0$.\footnote{\label{foo::nonexactcon} To be more precise, consistency with higher-spin symmetry \eqref{spinsgt} requires that $J_s$ has vanishing double-trace, and moreover is conserved up to pure trace terms,
\begin{equation}
\left(\partial_u \cdot \nabla\right) J_s\left(x,u\right) \;\approx\; {\cal O}\left(u^2\right).
\end{equation}
As we shall demonstrate explicitly in \S \tcb{\ref{subsubsec::offshellcubic}} (supplemented by \S \tcb{\ref{app:imprcur}}), improvement terms (which do not contribute to on-shell vertices) can be added to the $J_s$ such that it is exactly conserved.} For tree-level diagrams involving a single exchange, this inversion is independent of the off-shell gauge fixing of the exchanged field, since the exchanged field couples to on-shell external legs. In this context, the bulk-to-bulk propagator can be determined disregarding terms proportional to gradients \cite{Francia:2007qt,Francia:2008hd,Manvelyan:2008ks} -- both in the equation of motion and in the solution. To wit, one may solve\footnote{The symbol $\left\{\left\{\bullet\right\}\right\}$ indicates a double-traceless projection: 
\begin{align}
    (\partial_u\cdot \partial_u)^2\left\{\left\{f(u,x)\right\}\right\}=0, \quad \text{and} \quad \left\{\left\{f(u,x)\right\}\right\}= f(u,x)  \qquad \text{iff} \qquad (\partial_u\cdot \partial_u)^2 f(u,x)=0.
\end{align}}
\begin{align}
\label{FronsdaltensorDD}
&\left[(\Box_1- m^2)-u^2_1(\partial_{u_1}\cdot \partial_{u_1})\right]\Pi_{s}(x_1,u_1;x_2,u_2)\\ \nonumber
& \hspace*{6.5cm} = -(1-\frac{1}{4} \,u^2_1\, \partial_{u_1} \cdot \partial_{u_1})^{-1} \left\{\left\{\left(u_1\cdot u_2\right)^s\delta^{d+1}\left(x_{12}\right)\right\}\right\},
\end{align}
up to gradient terms. It is then sufficient to make an ansatz that is free from gradient terms, 
\begin{equation}
\label{eq:ManifestTrace}\Pi_{s}(x_1,u_1;x_2,u_2) = \sum^{\lfloor s/2 \rfloor}_{k=0} \int^{\infty}_{-\infty}  d\nu \; g_{k}\left(\nu\right) \left(u_1^2\right)^{k} \left(u_2^2\right)^{k} \Omega_{\nu,s-2k}(x_1,u_1;x_2,u_2).
\end{equation}
Plugging the ansatz into \eqref{FronsdaltensorDD} fixes the functions $g_{s,k}\left(\nu\right)$ \cite{Bekaert:2014cea}
\begin{align}
\notag
g_{s,0}\left(\nu\right)&=\frac{1}{(\tfrac{d}{2}+s-2)^2+\nu^2},\\
\label{answertrgauge}
g_{s,k}\left(\nu\right)&=-\frac{\left(1/2\right)_{k-1}}{2^{2k+3}\cdot k!}\frac{(s-2k+1)_{2k}}{(\tfrac{d}{2}+s-2k)_k (\tfrac{d}{2}+s-k-3/2)_k}\\
\notag
&\hspace*{2cm}\times \;
\frac{\left({(\tfrac{d}{2}+s-2k+i\nu)}/{2}\right)_{k-1}\left({(\tfrac{d}{2}+s-2k-i\nu)}/{2}\right)_{k-1}}{\left({(\tfrac{d}{2}+s-2k+1+i\nu)}/{2}\right)_{k}\left({(\tfrac{d}{2}+s-2k+1-i\nu)}/{2}\right)_{k}}, \qquad k\ne 0.
\end{align}
As for the massive propagators in the previous section, the $k=0$ term is the traceless and transverse part of the propagator which encodes the propagating degrees of freedom, while those for $k>0$ generate purely contact terms in exchange amplitudes.

\subsection{CPWE of spinning exchange diagrams}
\label{subsec::cpwesexch}
In this section we put together the results of the preceding sections to determine CPWEs of tree-level four-point exchange Witten diagrams with fields of arbitrary mass and integer spin on the internal and external legs.

\subsubsection{Natural basis of conformal partial waves in AdS/CFT}

To this end, it is useful to briefly discuss the integral form \eqref{mass123factor} of spinning conformal partial waves in terms of the natural AdS/CFT basis \eqref{nicebasis} of three point conformal structures. Employing this new basis, the spinning conformal partial waves of \S \tcb{\ref{subsubsec::cpinningcpwe}} read
\begin{align}\label{mass123factornewbasis}
& W^{\text{{\bf n}},\text{{\bf m}}}_{\Delta,s}\left(y_i\right) \; + \; \text{shadow} \\ \nonumber
& \hspace*{.75cm} = {\kappa}_{d-\Delta,s}\frac{\gamma_{\tau,s} {\bar \gamma}_{\tau,s}}{\pi^{d/2}} \int d^dy\, \left[\left[{\cal O}_{\Delta_1,s_1}(y_1){\cal O}_{\Delta_2,s_2}(y_2) {\cal O}_{\Delta,s}(y)  \right]\right]^{(\text{{\bf n}})} [[{\tilde {\cal O}}_{\Delta,s}(y)  {\cal O}_{\Delta_3,s_3}(y_3){\cal O}_{\Delta_4,s_4}(y_4)]]^{(\text{{\bf m}})},
\end{align}
where for convenience we repeat here the form of the basis elements \eqref{nicebasis}
{\small 
\begin{align}\label{nicebasis2}
&\left[\left[{\cal O}_{\Delta_1,s_1}(y_1){\cal O}_{\Delta_2,s_2}(y_2) {\cal O}_{\Delta_3,s_3}(y_3)\right]\right]^{(\text{{\bf n}})} \equiv\frac{{\sf H}_1^{n_1}{\sf H}_2^{n_2}{\sf H}_3^{n_3}}{(y_{12})^{\delta_{12}}(y_{23})^{\delta_{23}}(y_{31})^{\delta_{31}}} 
\left[\prod_{i=1}^32^{\tfrac{\delta_{(i+1)(i-1)} }{2}+n_i-1} \Gamma \left(\tfrac{\delta_{(i+1)(i-1)} }{2}+n_i\right)\right]
\\& \hspace*{2cm} \times\left[\prod_{i=1}^3 q_i^{\frac{1-n_i}{2}-\frac{\delta_{(i+1)(i-1)} }{4}} J_{(\delta_{(i+1)(i-1)}+2n_i -2)/2}\left(\sqrt{q_i}\right)\right]
\,{\sf Y}_1^{s_1-n_2-n_3}{\sf Y}_2^{s_2-n_3-n_1}{\sf Y}_3^{s_3-n_1-n_2} \nonumber
\end{align}}

 In combination with the basis \eqref{Ical} of bulk cubic vertices, once the harmonic function decomposition of a given spinning Witten diagram is known the choice of basis \eqref{mass123factornewbasis} of spinning conformal partial waves makes its CPWE follow almost automatically. We demonstrate this explicitly in the following section.

\subsubsection{Generic spinning exchange diagram}
\label{subsubsec::genspiexch}

We consider a generic tree-level four-point exchange of a spin-$s$ field of mass $m^2$ between fields of spin $s_i$ and mass {\small $m^2_i$}. This is depicted for the ${\sf s}$-channel below,
\begin{equation}\label{exchsch}
\includegraphics[scale=0.45]{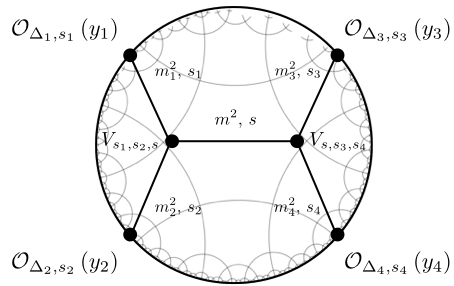}.
\end{equation}
At this level the cubic vertices $V_{s,s_i,s_j}$ are kinematic, and are not constrained by any consistency condition aside from the necessary requirement of respecting the AdS isometry. Expanded in the natural basis \eqref{Ical}, they read
\begin{equation}\label{gen3ptvertnice}
V_{s,s_i,s_j} = \sum_{n_i} g_{s_i,s_j,s}^{n_i,n_j,n}{\cal I}_{s_i,s_j,s}^{n_i,n_j,n},
\end{equation}
with arbitrary couplings $g_{s_i,s_j,s}^{n_i,n_j,n}$. 

The harmonic function decomposition decomposition follows upon insertion of the massive spin-$s$ bulk-to-bulk propagator \eqref{massiveans}
\begin{equation}\label{harmdecmpgen4pt}
 \includegraphics[scale=0.45]{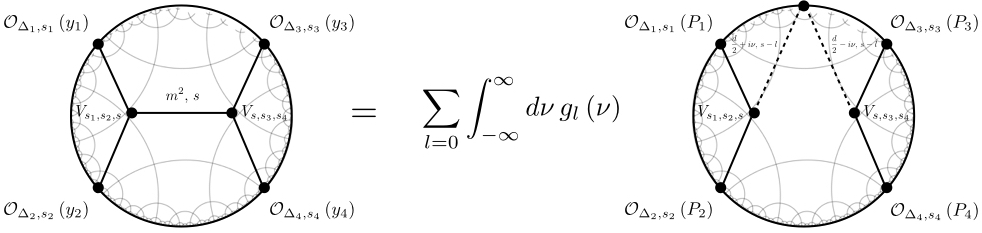}.
\end{equation}
What remains to determine the CPWE is to evaluate the three-point Witten diagrams on the RHS, which can be carried out seamlessly by employing the tools developed in \S \tcb{\ref{subsec::spinning3pt}}.

For this generic case we focus on the part of the exchange which encodes the propagating degrees of freedom. These are carried by the traceless and transverse part of the bulk-to-bulk propagator \eqref{massivett}, and accordingly we focus on the $l=0$ contribution in the harmonic function decomposition \eqref{harmdecmpgen4pt}. The latter is factorised into three-point Witten diagrams of the form:
\begin{equation}\label{im3ptgen}
\includegraphics[scale=0.3]{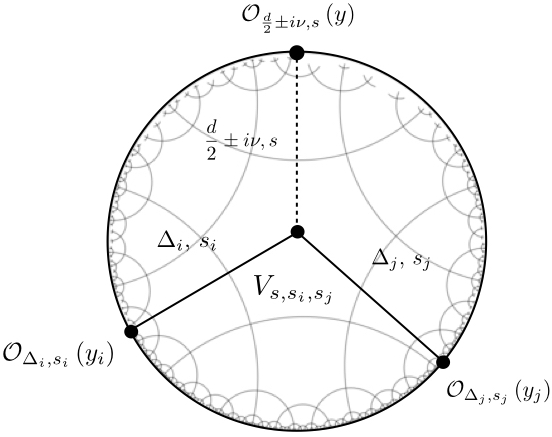},
\end{equation}
by virtue of the split representation \eqref{splitharm} of the harmonic function. The amplitudes \eqref{im3ptgen} can be straightforwardly given in any basis of three-point conformal structures using the results of \S \tcb{\ref{subsec::spinning3pt}}. However, choosing to expand the couplings \eqref{gen3ptvertnice} in the natural AdS/CFT basis \eqref{Ical} of cubic vertices gives the following simple and compact form 
\begin{align}\label{indi3ptmass}
 {\cal A}_{{\bf s};{\bm{\tau}^\pm}}\left(y_i,y_j,y\right) & = \sum_{{\bf n}} g_{\bf s}^{\bf n} {\cal A}^{{\bf n}}_{{\bf s};{\bm{\tau}^\pm}}\left(y_i,y_j,y\right) \\ \nonumber
& = \sum_{{\bf n}} g_{\bf s}^{\bf n}\, {\sf B}({\bf s};{\bf n};{\bm{\tau}^\pm})\,[[{\cal O}_{\Delta_i,s_i}(y_i){\cal O}_{\Delta_j,s_j}(y_j) {\cal O}_{\frac{d}{2}\pm i\nu,s}(y)]]^{(\text{{\bf n}})}
\end{align}
where we defined the vectors {\small ${\bf s} = \left(s_i,s_j,s\right)$}, {\small ${\bf n} = \left(n_i,n_j,n\right)$} and {\small $\bm{\tau}^\pm = (\tau_i,\tau_j,\tfrac{d}{2}\pm i\nu-s)$}.\footnote{In particular, $ g_{\bf s}^{\bf n} = g_{s_i,s_j,s}^{n_i,n_j,n}$ and ${\cal A}^{{\bf n}}_{{\bf s};{\bm{\tau}}}$ is the amplitude \eqref{IntegralBasisAdS} with labels ${\bf s} = \left(s_i,s_j,s\right)$, ${\bf n} = \left(n_i,n_j,n\right)$ and $\bm{\tau} = (\tau_i,\tau_j,\tfrac{d}{2}\pm i\nu-s)$.}

Using the integral representation of the conformal partial waves \eqref{mass123factornewbasis}, one can then immediately write down the CPWE of the ${\sf s}$-channel exchange \eqref{exchsch}\footnote{Here, ${\bf s}_{i,j} = \left(s_i,s_j,s\right)$, $\bm{\tau}^\pm_{i,j} = (\tau_i,\tau_j,\frac{d}{2}\pm i\nu-s)$, ${\bf n} = \left(n_1,n_2,n\right)$ and ${\bf m} = \left(m_3,m_4,m\right)$.} \footnote{The $...$ denote contact terms generated by the $l>0$ contributions in the harmonic function decomposition \eqref{harmdecmpgen4pt}.}
\begin{align}\nonumber
{\cal A}^{{\sf s}}_{s_1,s_2|s,m^2|s_3,s_4} & = \int^{\infty}_{-\infty} \frac{d\nu}{\nu^2+\left(\frac{d}{2}-\Delta\right)^2} \frac{\nu^2}{\pi} \,  {\cal A}_{{\bf s}_{1,2};{\bm{\tau}^+_{1,2}}}\left(y_1,y_2,y\right)  {\cal A}_{{\bf s}_{3,4};{\bm{\tau}^-_{3,4}}}\left(y_3,y_4,y\right)  + ... \\ \label{genercpwe1234}
& = \int^{\infty}_{-\infty} d\nu\, \sum_{{\bf n},{\bf m}} {\sf c}_{{\bf n},\textbf{m}}\left(\nu\right) W^{\text{{\bf n}},\text{{\bf m}}}_{\frac{d}{2}+i\nu,s}\left(y_i\right)+...\,.
\end{align}
with\footnote{To obtain this expression we used that \begin{align}
&  {\sf B}({\bf s}_{34};{\bf m};\bm{\tau}^-_{3,4}) = \frac{i\kappa_{d-i\nu,s} \gamma_{\frac{d}{2}+i\nu-s,s}{\bar \gamma}_{\frac{d}{2}+i\nu-s,s}}{2 \nu \pi^{d/2} C_{\frac{d}{2}+i\nu,s}} {\sf B}({\bf s}_{34};{\bf m};\bm{\tau}^+_{3,4}).
\end{align}}
\begin{align}
{\sf c}_{{\bf n},\textbf{m}}\left(\nu\right) = - g_{{\bf s}_{1,2}}^{{\bf n}}g_{{\bf s}_{3,4}}^{{\bf m}} \frac{{\sf B}({\bf s}_{12};{\bf n};\bm{\tau}^+_{1,2}){\sf B}({\bf s}_{34};{\bf m};\bm{\tau}^+_{3,4})}{\nu^2+\left(\frac{d}{2}-\Delta\right)^2} \frac{2\pi^{\tfrac{d}{2}-1}\Gamma\left(i\nu+1\right) \nu}{\left(i\nu+s+\frac{d}{2}-1\right)\Gamma\left(i\nu+\frac{d}{2}-1 \right)} .
\end{align}
This is the contour integral form \eqref{contcpwe} of the conformal partial wave expansion, reviewed in \S \tcb{\ref{subsubsec::extscalar}}. Recall that the functions ${\sf c}_{{\bf n},\textbf{m}}\left(\nu\right)$ encode the contribution from spin-$s$ operators: A pole at scaling dimension $\frac{d}{2}+i\nu = \lambda$ in the lower-half $\nu$-plane signifies a contribution from the conformal multiplet $\left[\lambda,s\right]$, whose residue gives the corresponding OPE coefficient. Separating out such poles into a function ${\sf p}_{{\bf n},\textbf{m}}\left(\nu\right)$, 
\begin{equation}
{\sf c}_{{\bf n},\textbf{m}}\left(\nu\right)={\bar {\sf c}}_{{\bf n},\textbf{m}}\left(\nu\right) {\sf p}_{{\bf n},\textbf{m}}\left(\nu\right),
\end{equation}
we have
{\small \begin{multline} \label{polemn}
{\sf p}_{{\bf n},\textbf{m}}\left(\nu\right) = \frac{1}{\nu^2+\left(\frac{d}{2}-\Delta\right)^2} \Gamma \left(\tfrac{2\left(s_1+n-n_2\right)+\tau_1+\tau_{2}+s-\left(\frac{d}{2}+i\nu\right)}{2}\right) \Gamma \left(\tfrac{2\left(s_2+n-n_1\right)+\tau_1+\tau_{2}+s-\left(\frac{d}{2}+i\nu\right)}{2}\right)    \\
\times  \Gamma \left(\tfrac{2\left(s_3+m-m_4\right)+\tau_3+\tau_{4}+s-\left(\frac{d}{2}+i\nu\right)}{2}\right) \Gamma \left(\tfrac{2\left(s_4+m-m_3\right)+\tau_3+\tau_{4}+s-\left(\frac{d}{2}+i\nu\right)}{2}\right). 
\end{multline}}

\noindent
There are two types of contributions, in accord with the standard lore on CPWEs of Witten diagrams \cite{Liu:1998ty,Liu:1998th,Freedman:1998bj,DHoker:1998ecp,DHoker:1998bqu,DHoker:1999kzh,Hoffmann:2000tr,Hoffmann:2000mx,Arutyunov:2000py,Arutyunov:2000ku,Hoffmann:2000tb,Uruchurtu:2007kq,ElShowk:2011ag,Sleight:2016hyl}:
\begin{enumerate}
\item {\bf Single-trace:} This is the universal contribution to an exchange diagram, corresponding to the exchange of the bulk single-particle state. Accordingly, it is generated by the pole-factor in the traceless and transverse part of the bulk-to-bulk propagator \eqref{massivett}, which carries the propagating degrees of freedom. This translates into a pole at $\frac{d}{2}+i\nu = \Delta$ in \eqref{polemn}, which coincides with the scaling dimension of the spin-$s$ single-trace operator ${\cal O}_{\Delta,s}$ that is dual to the exchanged spin-$s$ single-particle state of mass {\small $m^2R^2 = \Delta\left(\Delta-d\right)-s$} in the bulk.

\item {\bf Double-trace:} The remaining contributions originate from contact terms, arising from the collision of the two points that are integrated over the entire volume of AdS. This generates 2-particle states in the bulk, which are dual to double-trace operators on the conformal boundary. Accordingly, the corresponding poles are encoded in the factors \eqref{beeeeeee} arising from the integration over AdS. In the pole-function \eqref{polemn} these are the origin of the two sets of Gamma function poles ($p=0,1,2,3,...$)

{\small \begin{subequations}
\begin{align}
1. \quad \left(\frac{d}{2}+i\nu \right) -s & = \tau_1+\tau_{2}+2\left(s_1+n-n_2+p\right), \quad \left(\frac{d}{2}+i\nu \right) -s  = \tau_1+\tau_{2}+2\left(s_2+n-n_1+p\right)\\
2. \quad \left(\frac{d}{2}+i\nu \right) -s & = \tau_3+\tau_{4}+2\left(s_3+m-m_4+p\right), \quad \left(\frac{d}{2}+i\nu \right) -s = \tau_3+\tau_{4}+2\left(s_4+m-m_3+p\right),
\end{align}
\end{subequations}}

\nonumber
corresponding to contributions from the two families $\left[{\cal O}_{\Delta_1,s_1}{\cal O}_{\Delta_2,s_2}\right]_s$ and $\left[{\cal O}_{\Delta_3,s_3}{\cal O}_{\Delta_4,s_4}\right]_s$ of spin-$s$ double-trace operators, respectively. In the bulk, these correspond to 2-particle states created, respectively, by $\varphi_{s_1}$ with $\varphi_{s_2}$, and $\varphi_{s_3}$ with $\varphi_{s_4}$.

Let us briefly comment on the $l>0$ contributions to the harmonic function decomposition \eqref{harmdecmpgen4pt}. As explained earlier these are purely contact terms, and likewise generate double-trace contributions $\left[{\cal O}_{\Delta_1,s_1}{\cal O}_{\Delta_2,s_2}\right]_{s-l}$ and $\left[{\cal O}_{\Delta_3,s_3}{\cal O}_{\Delta_4,s_4}\right]_{s-l}$ to the CPWE, but of lower spin $s-l$.

\end{enumerate}

\subsection{Spinning exchanges in the type A higher-spin gauge theory}
\label{subsec::hsgauge}

So far our dialogue has not been restricted to any particular theory of spinning fields. In recent years, a lot of interest has been generated in theories of higher-spin gauge fields, owing in part to the conjectured duality \cite{HaggiMani:2000ru,Sundborg:2000wp,WittenTalk,Sezgin:2002rt,Klebanov:2002ja,Leigh:2003gk} between higher-spin gauge theories on AdS backgrounds and free CFTs. In this section we apply the tools and results of the preceding sections to compute all four-point exchange Witten diagrams in the simplest higher-spin gauge theory for $d>2$, which is known as the type A minimal higher-spin theory expanded about AdS$_{d+1}$ \cite{Vasiliev:2003ev}.\footnote{See \cite{Giombi:2009wh,Giombi:2010vg,Chang:2011mz,Ammon:2011ua,Hijano:2013fja,Bekaert:2014cea,Bekaert:2015tva,Sleight:2016dba,Sleight:2016hyl} for other results on Witten diagrams in higher-spin gauge theories.} This theory is conjectured to be dual to the (singlet sector of the) free scalar $O\left(N\right)$ model in $d$-dimensions \cite{Sezgin:2002rt,Klebanov:2002ja}.

The spectrum consists of a tower of totally symmetric even spin gauge fields (one for each even spin $s=2,4,6,...$) and a parity even scalar of mass $m^2_0R^2 = -2(d-2)$, which sits in the higher-spin multiplet. 

Before moving to the computation of the exchange amplitudes, we first review the result for the metric-like cubic couplings established in \cite{Sleight:2016dba}.
  
\subsubsection{Off-shell cubic couplings}
\label{subsubsec::offshellcubic} 

The off-shell cubic couplings of the type A minimal higher-spin theory on AdS$_{d+1}$ were determined in \cite{Sleight:2016dba}, for de Donder gauge.\footnote{Note that although the result \eqref{dedonder123} for the complete cubic couplings was fixed using the holographic duality, it was later verified \cite{Sleight:2016xqq} that the result solves the Noether procedure -- i.e. requiring that each cubic coupling is local, the cubic vertices coincide with those that would be obtained without employing holography. The result built upon the covariant classification \cite{Joung:2011ww,Joung:2012fv,Taronna:2012gb,Joung:2012hz,Joung:2013nma} of cubic interactions in AdS$_{d+1}$.} For tree-level exchanges we only require couplings with a single field -- the one that is exchanged -- off-shell. For a spin-$s$ field $\varphi_s$ in de Donder gauge, its interaction with two on-shell fields of spins $s_1$ and $s_2$ reads
\begin{align}\label{dedonder123}
&{\cal V}_{s_1,s_2,s}= g_{s_1,s_2,s}\left[1-\frac{1}{2}\left(d-2+{\cal Y}_i\partial_{{\cal Y}_i}\right)\partial^2_{{\cal Y}_3}\partial^2_{U_3}\right] {\cal Y}^{s_1}_1{\cal Y}^{s_2}_2{\cal Y}^{s}_3 \,\\ \nonumber
& \hspace*{7cm} \times \varphi_{s_1}\left(X_1,U_1\right)\varphi_{s_2}\left(X_2,U_2\right)\varphi_s\left(X_3,U_3\right).
\end{align}
The coupling constants $g_{s_1,s_2,s}$, for canonically normalised kinetic terms, are given by \cite{Sleight:2016dba}
\begin{equation}
g_{s_1,s_2,s} = \frac{1}{\sqrt{N}}\frac{\pi ^{\frac{d-3}{4}}2^{\tfrac{3 d-1+s_1+s_2+s}{2}}}{ \Gamma (d+s_1+s_2+s-3)} \sqrt{\frac{\Gamma(s+\tfrac{d-1}{2})}{\Gamma\left(s+1\right)}} \prod_{i=1}^2\sqrt{\frac{\Gamma(s_i+\tfrac{d-1}{2})}{\Gamma\left(s_i+1\right)}}.
\end{equation}
The first term in \eqref{dedonder123} is the traceless and transverse part of the vertex, which is non-trivial on-shell. The second term accounts for the off-shell de Donder field $\varphi_s$, and accordingly is proportional to its trace.

For the four-point exchange of a spin-$s$ gauge field, we massage the vertices \eqref{dedonder123} into the form 
\begin{equation}
{\cal V}_{s_1,s_2,s}\left(X\right) = s! J_{s|s_1,s_2}\left(X,\partial_U\right)\varphi_s\left(X,U\right), 
\end{equation}
with the spin-$s$ current $J_{s|s_1,s_2}$ bi-linear in $\varphi_{s_1}$ and $\varphi_{s_2}$. This is an exercise of integration by parts in ambient space, and gives
\begin{align} \label{j12}
J_{s|s_1,s_2} = \left(\sum_{k=0}^{\text{min}(s_1,s_2)}\tfrac{(-2)^k}{k!}\tfrac{\Gamma(s_1+s_2+s+d-3)}{\Gamma(s_1+s_2+s+d-3-k)}\tfrac{\Gamma(s_1+1)}{\Gamma(s_1-k+1)}\tfrac{\Gamma(s_2+1)}{\Gamma(s_2-k+1)}\mathcal{H}_3^k\,\bar{\mathcal{Y}}_1^{s_1-k}\bar{\mathcal{Y}}_2^{s_2-k}\bar{\mathcal{Y}}_3^{s}\right) + ...\,,
\end{align}
where the $...$ are terms that constitute the completion with $\varphi_s$ off shell, which are reinstated below. For convenience above we defined the contractions 
\begin{equation}
{\cal \bar Y}_1 = {\cal Y}_1, \qquad {\cal \bar Y}_2 = - \partial_{U_2} \cdot \partial_{X_1}, \qquad {\cal \bar Y}_3 = \tfrac{1}{2} \partial_{U_3} \cdot \left( \partial_{X_1}-\partial_{X_2} \right), \qquad {\cal H}_3 = \partial_{U_1} \cdot \partial_{U_2}.
\end{equation}
In its present form, the complete current \eqref{j12} is not exactly conserved. Indeed, recall that for a doubly-traceless Fronsdal field $\varphi_s$, higher-spin symmetry at the linearised level only requires that it is conserved up to traces (c.f. footnote \ref{foo::nonexactcon}). On the other hand, as emphasised in \S \tcb{\ref{subssec::masslessprop}}, the manifest trace form \eqref{eq:ManifestTrace} of the bulk-to-bulk propagators requires the use of exactly conserved currents. In appendix \S \tcb{\ref{app:imprcur}} we show the details of how the current \eqref{j12} can be improved such that it satisfies exact conservation. Here we just state that it can be attained by taking on-shell non-trivial part of \eqref{j12} and dressing each term with a differential operator
\begin{equation}
\mathcal{H}_3^k\,\bar{\mathcal{Y}}_1^{s_1-k}\bar{\mathcal{Y}}_2^{s_2-k}\bar{\mathcal{Y}}_3^{s} \: \rightarrow \: \left(\sum_{n=0}^{[s/2]}\alpha_{n}^{(k)}\,(\pl^2_{{\bar {\cal Y}}_3})^{n}(\pl_{U_3}^2)^n\right) \mathcal{H}_3^k\,\bar{\mathcal{Y}}_1^{s_1-k}\bar{\mathcal{Y}}_2^{s_2-k}\bar{\mathcal{Y}}_3^{s},
\end{equation}
where 
\begin{equation}
\alpha_{n}^{(k)}=\left(\frac{1}{2}\right)^{2n}\frac1{n!}\frac{\Gamma(3+k-s_1-s_2-\tfrac{d}{2}+n)}{\Gamma(3+k-s_1-s_2-\tfrac{d}{2})}\,.
\end{equation}

\subsubsection{Four-point exchange diagrams}

Consider the four-point exchange of a spin-$s$ gauge field between gauge fields of spin $s_i$ in the ${\sf s}$-channel. The manifest trace form of the bulk-to-bulk propagator \eqref{eq:ManifestTrace} gives the harmonic function decomposition
\begin{equation}\label{exchfigh}
\includegraphics[scale=0.4]{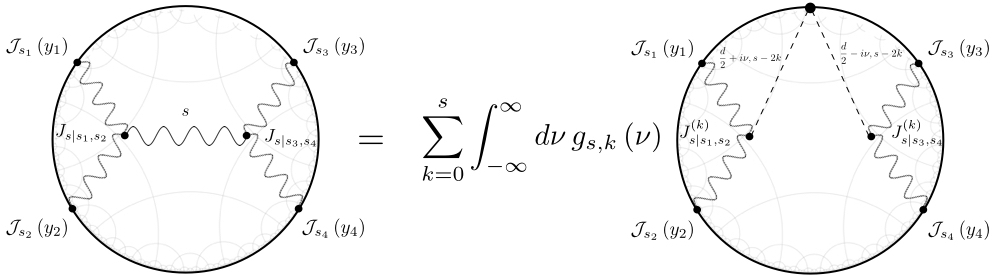},
\end{equation}
where the operator ${\cal J}_{s_i}$ is the spin-$s_i$ conserved current in the free scalar $O\left(N\right)$ model dual to the spin-$s_i$ gauge field $\varphi_{s_i}$ in the bulk. The notation $J^{(k)}$ denotes the $k$-th trace of the conserved current $J$, which arise from the trace structure of the bulk-to-bulk propagator contact terms. The explicit form of $J$ is given in \S \tcb{\ref{subsubsec::offshellcubic}}, while its $k$-th trace is derived in \S \tcb{\ref{app::trofcur}}.

To determine the CPWE, we therefore need to evaluate three-point Witten diagrams of the form, 
\begin{equation}\label{fig3ptk}
\includegraphics[scale=0.325]{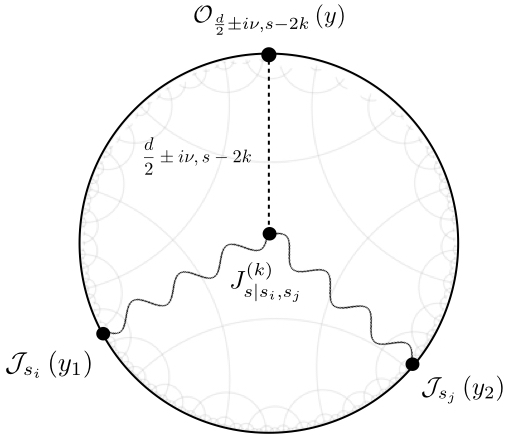},
\end{equation}
which, employing the tools introduced in \S \tcb{\ref{subsec::spinning3pt}} entails expressing the cubic couplings in the basis \eqref{Ical}.

We focus first on the $k=0$ contribution, which encodes the propagating degrees of freedom. As we saw for the massive exchanges in \S \tcb{\ref{subsubsec::genspiexch}}, this is generated by the traceless and transverse part of the bulk-to-bulk propagator \eqref{eq:ManifestTrace}. Accordingly, only the on-shell non-trivial (traceless and transverse) part of the cubic couplings 
\eqref{dedonder123} contribute, whose explicit form we give here for convenience:
\begin{equation}\label{TT123}
{\cal V}^{TT}_{s_1,s_2,s}\left(X\right)= g_{s_1,s_2,s}{\cal Y}^{s_1}_1{\cal Y}^{s_2}_2{\cal Y}^{s}_3 \varphi_{s_1}\left(X_1,U_1\right)\varphi_{s_2}\left(X_2,U_2\right)\varphi_s\left(X_3,U_3\right)\Big|_{X_i=X}.
\end{equation}
Nicely, this is already in the natural AdS/CFT basis \eqref{Ical} and the amplitudes \eqref{fig3ptk} for $k=0$ can be immediately written down by employing the result \eqref{IntegralBasisAdS}
\begin{align}\label{bulk3ptstr}
{\cal A}_{{\bf s};\bm{\tau}^\pm}\left(y_i,y_j,y\right) = g_{s_i,s_j,s}\,{\sf B}\left({\bf s};{\bf 0};\bm{\tau}\right)[[{\cal J}_{s_i}\left(y_i\right){\cal J}_{s_j}\left(y_j\right){\cal O}_{\tfrac{d}{2}\pm i\nu,s}\left(y\right)]]^{\bm{0}},
\end{align}
where here $\bm{\tau}^\pm=\left(d-2,d-2,\tfrac{d}{2}\pm i\nu-s\right)$ and ${\bf n}=(0,0,0)$. 

Following the discussion of \S \tcb{\ref{subsubsec::genspiexch}} for the generic case, one then obtains that the $k=0$ term in the harmonic function decomposition \eqref{exchfigh} of the exchange diagram yields the following contributions to its CPWE  
\begin{align}
   {\cal A}^{{\sf s}}_{s_1,s_2|s|s_3,s_4} = \int^\infty_{-\infty} d\nu\, {\sf c}_{s}\left(\nu\right)  W^{{\bf 0},{\bf 0}}_{\frac{d}{2}+i\nu,s}\left(y_i\right) + ...
\end{align}
with
\begin{align}
{\sf c}_{s}\left(\nu\right) = - g_{s_1,s_2,s}g_{s_3,s_4,s} \frac{{\sf B}({\bf s}_{12};{\bf 0};\bm{\tau}^+){\sf B}({\bf s}_{34};{\bf 0};\bm{\tau}^+)}{\nu^2+\left(s+\frac{d}{2}-2\right)^2} \frac{2\pi^{\tfrac{d}{2}-1}\Gamma\left(i\nu+1\right) \nu}{\left(i\nu+s+\frac{d}{2}-1\right)\Gamma\left(i\nu+\frac{d}{2}-1\right)},
\end{align}
where the pole function \eqref{polemn} in this case is given by 
\begin{multline}
\label{typeapole}
   {\sf p}_s\left(\nu\right) = \frac{1}{\nu^2+\left(s+\frac{d}{2}-2\right)^2} \Gamma \left(\tfrac{2\left(s_1+d-2\right)+s-\left(\frac{d}{2}+i\nu\right)}{2}\right) \Gamma \left(\tfrac{2\left(s_2+d-2\right)+s-\left(\frac{d}{2}+i\nu\right)}{2}\right)    \\
\times  \Gamma \left(\tfrac{2\left(s_3+d-2\right)+s-\left(\frac{d}{2}+i\nu\right)}{2}\right) \Gamma \left(\tfrac{2\left(s_4+d-2\right)+s-\left(\frac{d}{2}+i\nu\right)}{2}\right).
\end{multline}
In the following we discuss in detail the particular contributions.

\subsubsection*{Single-trace}

In line with the discussion of the generic case in \S \tcb{\ref{subsubsec::genspiexch}}, the pole factor in the traceless and transverse part of the bulk-to-bulk propagator \eqref{eq:ManifestTrace} generates a pole in \eqref{typeapole} at {\small $\frac{d}{2}+i\nu = s+d-2$}, which is the scaling dimension of the dual spin-$s$ conserved current ${\cal J}_s$ in the free scalar $O\left(N\right)$ model.

Furthermore, notice that for {\small $\frac{d}{2}+i\nu = s+d-2$} the three-point conformal structure generated by the $k=0$ amplitude \eqref{bulk3ptstr} coincides with the three-point conserved structure \eqref{besslefon} in free scalar theories. The corresponding spin-$s$ conformal partial wave thus coincides with the conserved conformal partial wave {\small ${\cal W}^{0,0}_{s_1,s_2|s|s_3,s_4}$} in the set \eqref{consshad}, which represents the contribution from the conserved operator {\small ${\cal J}_s$} to the four-point function {\small $\langle{\cal J}_{s_1}{\cal J}_{s_2}{\cal J}_{s_3}{\cal J}_{s_4}\rangle$} in free scalar theories.

With the result \eqref{onopecoef} of all single-trace conserved current OPE coefficients in free scalar theories, in this case we can confirm the standard expectation that the single-trace contribution to an exchange Witten diagram \emph{coincides} with the contribution from the same single-trace operator in the CPWE of the dual CFT four-point function, when expanded in the same channel. Indeed, using that \cite{Sleight:2016dba,Sleight:2016xqq}\footnote{Here we divide by the normalisation of the bulk-to-boundary propagators \eqref{buboprops} to give unit normalisation to the dual single-trace operator two point functions \eqref{2ptnorm}.}
\begin{equation}\label{opecoup}
g_{s_i,s_j,s}\frac{{\sf B}({\bf s}_{ij};{\bf 0};d-2,d-2,d-2)}{\sqrt{C_{s_i+d-2,s_i}C_{s_j+d-2,s_j}C_{s+d-2,s}}} = {\sf c}_{{\cal J}_{s_i}{\cal J}_{s_j}{\cal J}_{s}},
\end{equation}
we have (closing the contour in the lower-half $\nu$-plane)
\begin{equation}
-2 \pi i \text{Res}\,\left[{\sf c}_s\left(\nu\right) W^{{\bf 0},{\bf 0}}_{\frac{d}{2}+i\nu,s}\left(y_i\right),\tfrac{d}{2}+i\nu=s+d-2\right] = {\sf c}_{{\cal J}_{s_1}{\cal J}_{s_2}{\cal J}_{s}}{\sf c}^{{\cal J}_{s}}{}_{{\cal J}_{s_3}{\cal J}_{s_4}}{\cal W}^{0,0}_{s_1,s_2|s|s_3,s_4}\left(y_i\right),
\end{equation}
as expected.

\subsubsection*{Double-trace}

In this case the two sets of Gamma function poles 
{\small \begin{subequations}
\begin{align}
1. \quad \left(\frac{d}{2}+i\nu \right) -s & = 2\left(d-2\right)+2\left(p+s_1\right), \quad \left(\frac{d}{2}+i\nu \right) -s  = 2\left(d-2\right)+2\left(p+s_2\right),\\
2. \quad \left(\frac{d}{2}+i\nu \right) -s & = 2\left(d-2\right)+2\left(p+s_3\right), \quad \left(\frac{d}{2}+i\nu \right) -s = 2\left(d-2\right)+2\left(p+s_4\right),
\end{align}
\end{subequations}}
with $p=0,1,2,3,...\,$, correspond to contributions from the two families {\small $\left[{\cal J}_{s_1}{\cal J}_{s_2}\right]_s$} and {\small $\left[{\cal J}_{s_3}{\cal J}_{s_4}\right]_s$} of spin-$s$ double-trace operators build from single-trace conserved currents.

\subsubsection*{$k>0$ contributions}

Similarly, being contact, the $k>0$ contributions to the CPWE of the exchange \eqref{exchfigh} are from double-trace operators {\small $\left[{\cal J}_{s_1}{\cal J}_{s_2}\right]_{s-2k}$} and {\small $\left[{\cal J}_{s_3}{\cal J}_{s_4}\right]_{s-2k}$} of lower spin $s-2k$. Computing the corresponding three-point Witten diagrams \eqref{fig3ptk} for $k>0$ is a lot more involved, as it requires to compute the $k$-th trace of the currents {\small $J^{(k)}_{s|s_i,s_j}$}. We give a recipe for computing them in \S \tcb{\ref{app::trofcur}}, but stop short of evaluating the corresponding Witten diagrams; given that the $s_1$-$s_2$-$s_3$-$s_4$ quartic contact vertex is currently unfixed in metric-like form for the type A minimal higher-spin theory, the $k>0$ contributions are anyway highly dependent on the choice of field frame.

We non-the-less point out that there are simplifications for particular combinations of the external spins, such as for a single spinning external field (e.g. $s_1$-$0$-$0$-$0$) and also for a single spinning external field either side of the exchange (e.g. $s_1$-$0$-$s_3$-$0$ in the {\sf s}-channel), where the three-point bulk integrals for $k>0$ are of the same type as in the $k=0$ case.

\subsection*{Acknowledgements}

C. S. and M. T. are grateful to M.~Henneaux for useful discussions, and also A.~Castro, E.~Llabr\'es and F. G.~Rej\'on-Barrera in the context of geodesic Witten diagrams. C. S. also thanks D.~Francia for useful correspondence. The research of M. T. is partially supported by the Fund for Scientific Research-FNRS Belgium, grant FC 6369 and by the Russian Science Foundation grant 14-42-00047 in association with Lebedev Physical Institute.

\appendix

\section{Conventions, notations and ambient space}
\label{app::cna}

In this work we employ the same conventions as in \cite{Sleight:2016dba}, which we very briefly review here for completeness. For more details on the ambient space formalism, see for instance \cite{Grigoriev:2011gp,Taronna:2012gb,Sleight:2017krf}.

The ambient formalism is an indispensable framework for computations in AdS$_{d+1}$ space. In this context, the latter is viewed as a hyperboloid embedded in an ambient $\left(d+2\right)$-dimensional Minkowski space
\begin{align}
    X^2+R^2&=0\,, \qquad X^0>0\,, \label{adsamb}
\end{align}
where $R$ is the AdS radius. In ambient light-cone coordinates $(X^+, X^-, X^i)$ with $X^2=-X^+ X^- +\delta_{ij}X^i X^j$, the solution of the constraints \eqref{adsamb} in the Poincar\'e co-ordinates $x^\mu = \left(z,y^i\right)$ is given by
\begin{equation}\label{poin}
X^A=\frac{R}{z} (1,z^2+y^2,y^i)\,.
\end{equation}

\subsubsection*{Bulk fields}

In order to obtain a one-to-one correspondence between fields on AdS and those living in the higher-dimensional flat ambient space, one imposes constraints with defining the ambient space extensions of the AdS$_{d+1}$ fields \cite{Fronsdal:1978vb}. Such restrictions are usually given as homogeneity and tangentiality constraints. Employing a generating function formalism with intrinsic and ambient auxiliary vectors $u^\mu$ and $U^A$, a symmetric rank-$s$ tensor $\varphi_s\left(x,u\right)$ intrinsic to the AdS manifold is represented in ambient space by 
\begin{equation}
 \varphi_s\left(x,u\right) = \frac{1}{s!} \varphi_{\mu_1 ... \mu_s}(x)u^{\mu_1}...u^{\mu_s} \quad \rightarrow \quad \varphi_s(X,U)=\frac1{s!}\,\varphi_{A_1\ldots A_s}(X)U^{A_1}\ldots U^{A_s}\,.
\end{equation}
subject to the following homogeneity and tangentiality conditions
\begin{align}\label{tanhomo}
    (X\cdot\pl_X-\Delta)\varphi_s(X,U)&=0\,,& \left(X\cdot\pl_U\right) \varphi_s(X,U)&=0\,.
\end{align}
Nicely, the conditions \eqref{tanhomo} imply that on-shell
\begin{equation}
\partial^2_X \varphi_{A_1\ldots A_s} = 0,
\end{equation}
for the ambient representative of the AdS field $\varphi_s$ of mass $m^2R^2=\Delta\left(\Delta-d\right)-s$.

Let us stress that in imposing tangentiality and homogeneity conditions \eqref{tanhomo} one is implicitly extending the AdS field to the full ambient space, where $X^2$ plays the role of the radial coordinate. This formalism is different from the manifestly intrinsic formalism (for instance used in \cite{Costa:2014kfa}) where one never moves away from the AdS manifold $X^2=-R^2$.

The ambient representative of the AdS covariant derivative $\nabla_{\mu}$ takes the simple form 
\begin{equation}
    \nabla_{A} = {\cal P}^{B}_{A}\frac{\partial}{\partial X^B},\label{ambcovd}
\end{equation}
and acts via
\begin{equation}
\nabla = {\cal P} \circ \partial \circ {\cal P}.
\end{equation}

\subsubsection*{Boundary fields}

The boundary of AdS$_{d+1}$ is identified with the null rays
\begin{equation}\label{nullcon}
P^2=0, \qquad P\:\sim\: \lambda P, \qquad \lambda \ne 0,
\end{equation}
where $P$ gives the ambient space embedding of the CFT coordinate $y^i$. It is convenient to introduce the boundary analog of the auxiliary variables $U^A$, which we refer to as $Z_A(y)$ and extend to ambient space the null CFT auxiliary variable $z^i$. Working in light cone coordinates $P^A=(P^+,P^-,P^i)$, with the gauge choice $P^+=1$ one has
\begin{align}
     P^A(y)=(1,y^2,y^i)\, \qquad \text{and} \qquad Z^A(y)=(0,2y\cdot z,z^i)\,. \label{ps}
\end{align}
A symmetric rank-$s$ boundary operator $ O_{\Delta,s}$ of scaling dimension $\Delta$ is represented by:\footnote{Note that here $z$ denotes the auxiliary vector $z^i$ and should not be confused with the radial Poincar\'e co-ordinate in \eqref{poin}.}
\begin{equation}
    O_{\Delta,s}(y,z)=\frac1{s!}O_{\mu_1\ldots\mu_s}\left(y\right)\,z^{\mu_1}\cdots z^{\mu_s} \quad \rightarrow \quad \mathcal{O}_{\Delta,s}(P,Z)=\frac1{s!}\mathcal{O}_{A_1... A_s}\left(P\right)Z^{A_1}\cdots Z^{A_s}\,,
\end{equation}
where 
\begin{align}
(P\cdot\pl_P-\Delta)O_{\Delta,s}(P,Z)&=0\,,& \left(P\cdot\pl_Z\right) \mathcal{O}_{\Delta,s}(P,Z)&=0\,,
\end{align}
and, being restricted to the null cone \eqref{nullcon}, there is an extra redundancy
\begin{align}\label{extred}
 & \hspace*{3.5cm} O_{A_1...A_s}(P) \rightarrow O_{A_1...A_s}(P) + P_{\left(A_1\right.}\Lambda_{\left. A_2 ... A_s \right)},\\
 & P^{A_1}\Lambda_{ A_1 ... A_{s-1}} = 0, \quad \Lambda_{ A_1 ... A_{s-1}}(\lambda P) = \lambda^{-(\Delta+1)}\Lambda_{ A_1 ... A_{s-1}}(P), \quad \eta^{A_1A_2}\Lambda_{A_1 ... A_{s-1}} = 0.
 \end{align}

\section{The improved current}
\label{app:imprcur}
In this appendix we detail the improvement of the higher-spin currents \eqref{j12} to make them exactly conserved. We begin with the traceless and transverse part of the current,
\begin{align} \label{j12a}
J^{TT}_{s_3|s_1,s_2} = \sum_{k=0}^{\text{min}(s_1,s_2)}\tfrac{(-2)^k}{k!}\tfrac{\Gamma(s_1+s_2+s_3+d-3)}{\Gamma(s_1+s_2+s_3+d-3-k)}\tfrac{\Gamma(s_1+1)}{\Gamma(s_1-k+1)}\tfrac{\Gamma(s_2+1)}{\Gamma(s_2-k+1)}\mathcal{H}_3^k\,\bar{\mathcal{Y}}_1^{s_1-k}\bar{\mathcal{Y}}_2^{s_2-k}\bar{\mathcal{Y}}_3^{s_3}.
\end{align}
On-shell, each monomial in the above is conserved. One can therefore study the structure of the required improvements with $\varphi_3$ off-shell for a given monomial
\begin{equation}\label{monom123}
f_{s_1,s_2,s_3}^{(k)}=\mathcal{H}_3^k\,\bar{\mathcal{Y}}_1^{s_1-k}\bar{\mathcal{Y}}_2^{s_2-k}\bar{\mathcal{Y}}_3^{s_3}\,.
\end{equation}
The combination of different monomials in \eqref{j12a} above is necessary to achieve on-shell gauge invariance with respect to $\varphi_1$ and $\varphi_2$ which can be easily verified explicitly (see e.g. \cite{Taronna:2012gb}).

In order to proceed to find the conserved improvement, the doubly-traceless condition on $\varphi_3$ together with the traceless condition on the corresponding gauge parameter needs to be dropped. Not doing so would only recover a current whose traceless part is conserved. We hence consider the following ansatz for the improvement, dressing each monomial \eqref{monom123} with trace operators
\begin{equation}\label{ansfk}
F_{s_1,s_2,s_3}^{(k)}=\left(\sum_{n=0}^{[s_3/2]}\alpha_{n}^{(k)}\,(\pl^2_{{\bar {\cal Y}}_3})^{n}(\pl_{U_3}^2)^n\right)f_{s_1,s_2,s_3}^{(k)}\,,
\end{equation}
where the derivative with respect to ${\bar {\cal Y}}_3$ accounts for the fact that taking the trace lowers the spin. The coefficients $\alpha_{n}^{(k)}$ in the ansatz \eqref{ansfk} are fixed by requiring gauge invariance of the vertex with $\varphi_3$ off-shell and with traceless gauge parameter $\xi_3$
\begin{equation}
\int dX\, \left(U_3\cdot\nabla_3 \, \xi_3\right) F_{s_1,s_2,s_3}^{(k)} \varphi_1\varphi_2=0\,.
\end{equation}
In the above the fields $\varphi_1$ and $\varphi_2$ are on-shell, while the integral sign (which in the following will be omitted for ease of notation) implies that the above identity holds modulo total derivatives. 

Employing the explicit form of the gradient operator:
\begin{equation}
U_3\cdot\nabla_3=U_3\cdot\pl_{X_3}-\frac{U_3\cdot X_3}{X_3^2}\,(X_3\cdot\pl_{X_3}-U_3\cdot\pl_{U_3})\,,
\end{equation}
we arrive to the following conservation condition:
\begin{equation}
\sum_{n=0}^\infty\left[2(n+1)\alpha_{n+1}^{(k)}-\tfrac14\,\alpha_{n}^{(k)}[d+2(s_1+s_2-k-n)-4]\right](\pl^2_{{\bar {\cal Y}}_3})^{n}f_{s_1,s_2,s_3}^{(k)}(\pl_{U_3}^2)^n\pl_{U_3}\cdot\pl_{X_3}=0\,,
\end{equation}
which leads to the solution for $\alpha_{n}^{(k)}$ in the form
\begin{equation}
\alpha_{n}^{(k)}=\left(\frac{1}{2}\right)^{2n}\frac1{n!}\frac{\Gamma(3+k-s_1-s_2-\tfrac{d}{2}+n)}{\Gamma(3+k-s_1-s_2-\tfrac{d}{2})}\,.
\end{equation}

\section{Trace of the currents}
\label{app::trofcur}
In order to evaluate the current exchange \eqref{exchfigh} with the manifest trace form of the propagator \eqref{eq:ManifestTrace}, we are required to compute the $n$-th trace of the exactly conserved current derived in the previous section. The process can be simplified by noting that in the present context the traces are contracted with harmonic functions, where we encounter terms of the form
\begin{equation}\label{trtermsapp}
(\pl^2_{{\bar {\cal Y}}_3})^{n}f_{s_1,s_2,s_3}^{(k)}(\pl_{U_3}^2)^n(u_3^2)^q\,\Omega_{\nu,s_3-2q}\Big|_{U_3=0}\,,
\end{equation}
where $u_3^2$ is the intrinsic symmetrised metric tensor written in generating function form, which can be re-expressed in the ambient formalism as
\begin{equation}
u_3^2=U_3^2-\frac{U_3\cdot X_3}{X_3^2}\,.
\end{equation}

To evaluate the trace one commutes the $U_3$ contained in the $\left(u^2_3\right)^q$ to the far left hand side, where the condition $U_3=0$ can be applied. We first commute the $u_3^2$ past the $\pl_{U_3}^2$, which, employing the tracelessness of the harmonic functions, reads
\begin{align}
(\pl_{U_3}^2)^n(u_3^2)^q\Omega_{\nu,s_3-2q}&=A_n^q(u_3^2)^{q-n}\Omega_{\nu,s_3-2q}\,,& A_n^q&=2^{2 n}\,\tfrac{\Gamma (n-q)}{\Gamma (-q)}\tfrac{\Gamma \left(-\frac{d}{2}+n+q-s_3+\frac{1}{2}\right)}{\Gamma \left(-\frac{d}{2}+q-s_3+\frac{1}{2}\right)}\,.
\end{align}
What remains is to evaluate terms of the form
\begin{equation}\label{finalcommtr}
(\pl^2_{{\bar {\cal Y}}_3})^{n}f_{s_1,s_2,s_3}^{(k)}(u_3^2)^{q-n}\Omega_{\nu,s_3-2q}\Big|_{U_3=0}.
\end{equation}

To this end, it is useful split ${\bar {\cal Y}}_3$ as\footnote{Note that in fact $V_{31} =  {\cal Y}_3$, but for ease of notation in this section we employ the labelling $V_{31}$.}
\begin{equation}
{\bar {\cal Y}}_3 = \frac{1}{2} \left( V_{31} - V_{32} \right), \qquad  V_{31} = \partial_{U_3} \cdot \partial_{X_1}, \qquad V_{31} = \partial_{U_3} \cdot \partial_{X_2},
\end{equation}
and express any function of ${\bar {\cal Y}}_3$ instead in terms of $V_{31}$ and $V_{21}$. In this way, the action of some operator $g\left({\bar {\cal Y}}_3\right)$ on $u_3$ can be expressed in the form
\begin{equation}
g\left({\bar {\cal Y}}_3\right) u^A_3 \Big|_{U_3=0} = \left(\pl_{X_1}^A\pl_{V_{31}}+\pl_{X_2}^A\pl_{V_{32}}\right)g\left(V_{31},V_{32}\right),
\end{equation}
from which follows the general formula
\begin{multline}
g\left({\bar {\cal Y}}_3\right) u_3^2  \Big|_{U_3=0}  =\Big\{2\left[\pl_{X_1}\cdot\pl_{X_2}+X_1\cdot\pl_{X_1}\,X_2\cdot\pl_{X_2}\right]\pl_{V_{31}}\pl_{V_{32}}\\+X_1\cdot\pl_{X_1}(X_1\cdot\pl_{X_1}-1)\pl_{V_{31}}^2+X_2\cdot\pl_{X_2}(X_2\cdot\pl_{X_2}-1)\pl_{V_{32}}^2\Big\}g\left(V_{31},V_{32}\right)\,,
\end{multline}
which can be iteratively applied to evaluate the traces in \eqref{finalcommtr}.
Now, since each current in the exchange is to be integrated over AdS, we can evaluate the above terms up to integrations by parts using
\begin{align}
X_1\cdot\pl_{X_1}&=-(d-2+\bar{\mathcal{Y}}_1\pl_{\bar{\mathcal{Y}}_1}+\bar{\mathcal{Y}}_2\pl_{\bar{\mathcal{Y}}_2}+V_{31}\pl_{V_{31}}+Q_3\pl_{Q_3})\,,\\
X_2\cdot\pl_{X_2}&=-(d-2+\bar{\mathcal{Y}}_1\pl_{\bar{\mathcal{Y}}_1}+\bar{\mathcal{Y}}_2\pl_{\bar{\mathcal{Y}}_2}+V_{32}\pl_{V_{32}}+Q_3\pl_{Q_3}),
\end{align}
where 
\begin{align}
Q_3&=-\frac12(X_1\cdot\pl_{X_1}+X_1\cdot\pl_{X_1}+\Delta_3+d)(X_1\cdot\pl_{X_1}+X_1\cdot\pl_{X_1}-\Delta_3),
\end{align}
and $\Delta_3 = \frac{d}{2} \pm i \nu$.

After evaluating the action of the above operators one can integrate by parts to obtain an expression for the final form of the trace terms \eqref{trtermsapp} in the form
\begin{equation}
J^{\left(k\right)}_{s_3|s_1,s_2} \cdot \Pi_{\frac{d}{2} \pm i\nu,s_3-2k} = \sum^{\text{min}\left(s_1,s_2\right)}_{m=0} \beta^{k,m}_{s_3|s_1,s_2}  {\cal H}^{m}_3 {\cal Y}^{s_1-m}_1 {\cal Y}^{s_2-m}_2 {\cal Y}^{s_3-2k}_3,
\end{equation}
where, via integration by parts, we replaced $\bar{\mathcal{Y}}_2 \rightarrow e^{\lambda \mathcal{H}_3\pl_{\mathcal{Y}_1}\pl_{\mathcal{Y}_2}}$, $V_{31} \rightarrow \mathcal{Y}_3$ and $V_{32}\rightarrow -\mathcal{Y}_3 + \partial_{U_3} \cdot \partial_{X_3} = - \mathcal{Y}_3$, where in the latter equality we used that the harmonic function is divergenceless. See \cite{Taronna:2012gb,Sleight:2016dba} for (in-context) reviews of integration by parts in the ambient space formalism, where in particular the parameter $\lambda$ and its use is defined. 

The explicit form of the coefficients $\beta^{k,m}_{s_3|s_1,s_2}$ is rather involved, and since the results of this work do not rely on the knowledge of the explicit expression for contact terms in exchange amplitudes (which are anyway highly dependent on the field frame), we do not present them here.

\section{Seed bulk integrals}
\label{sec::seedint}

Our approach to evaluate spinning three-point Witten diagrams is underpinned by their differential relationship \eqref{iter} with basic seed diagrams with external scalars \cite{Freedman:1998tz}. The latter is the basic ingredient from which our results are generated, which we briefly review here. 

It is useful to employ the Schwinger-parameterised form for the propagator \cite{Penedones:2010ue,Paulos:2011ie}
\begin{align}\label{schwing}
K_{\Delta}\left(X;P\right) = \frac{C_{\Delta,0}}{\Gamma\left(\Delta\right)} \int^\infty_0 \frac{dt}{t} t^\Delta e^{2t P \cdot X},
\end{align}
which results in
\begin{align}
A^{0,0,0}_{0,0,0;\tau_1,\tau_2,\tau_3}\left(P_1,P_2,P_3\right) = \int^{\infty}_0 \prod\limits^{3}_{i=1}\left(\frac{C_{\Delta_i,0}}{\Gamma\left(\Delta_i\right)}\frac{dt_i}{t_i} t^{\Delta_i}\right) \int_{\text{AdS}} dX e^{2\left(t_1 P_1+t_2 P_2+t_3 P_3\right) \cdot X}.
\end{align}
The integration over AdS is then straightforward to perform, and yields (see e.g. box 5.2 in \cite{Sleight:2017krf})
\begin{align}\label{3ptcont}
&\int^{\infty}_0 \prod\limits^{3}_{i=1}\left(\frac{dt_i}{t_i} t^{\Delta_i}\right) \int_{\text{AdS}} dX e^{2\left(t_1 P_1+t_2 P_2+t_3 P_3\right) \cdot X} \\ \nonumber
&  \hspace*{3cm} = \pi^{\frac{d}{2}}\Gamma\left(\frac{- d + \sum\nolimits^3_{i=1} \Delta_i}{2}\right)  \int^{\infty}_0 \prod\limits^{3}_{i=1}\left(\frac{dt_i}{t_i} t^{\Delta_i}_i\right) e^{\left(-t_1t_2 P_{12} - t_1t_3 P_{13}-t_2t_3 P_{23}\right)},
\end{align}
where $P_{ij} = -2 P_i \cdot P_j$. Through the change of variables,
\begin{equation}
    t_1 = \sqrt{\frac{m_2m_3}{m_1}}, \quad t_2 = \sqrt{\frac{m_1m_3}{m_2}}, \quad t_3 = \sqrt{\frac{m_1m_2}{m_3}}, \label{tov}
\end{equation}
we then obtain the final result:
\begin{multline}
\label{123scal}
 A^{0,0,0}_{0,0,0;\tau_1,\tau_2,\tau_3}\left(P_1,P_2,P_3\right) \\ = \frac{\pi^{\frac{d}{2}}}{2}\,\Gamma\left(\frac{- d + \sum\nolimits^3_{i=1} \Delta_i}{2}\right) \left(\prod\limits^{3}_{i=1} \frac{C_{\Delta_i,0}}{\Gamma\left(\Delta_i\right)} \right)\int^{\infty}_0 \prod\limits^{3}_{i=1}\left(\frac{dm_i}{m_i} m_i^{\delta_{(i+1)(i-1)}/2}\right) \exp\left(-m_i P_{(i+1)(i-1)}\right) \,,
\end{multline}
where $i \cong i+3$ and 
\begin{equation}
\delta_{i(i+1)}=\frac{\Delta_i+\Delta_{(i+1)}-\Delta_{(i-1)}}2\,,
\end{equation}
 The standard three-point conformal structure for scalar operators is obtained from \eqref{123scal} using the integral representation of the Gamma function
\begin{equation}
\label{123scalnom}
 A^{0,0,0}_{0,0,0;\tau_1,\tau_2,\tau_3}\left(P_1,P_2,P_3\right) = {\sf C}\left(\Delta_1,\Delta_2,\Delta_3; 0\right)  \frac{1}{P^{\frac{\Delta_1+\Delta_3-\Delta_2}{2}}_{13}P^{\frac{\Delta_2+\Delta_3-\Delta_1}{2}}_{23}P^{\frac{\Delta_1+\Delta_2-\Delta_3}{2}}_{12}} \,,
\end{equation}
where explicitly 
\begin{align}\label{ofactorst}
   & {\sf C}\left(\Delta_1,\Delta_2,\Delta_3; 0\right) \\ \nonumber
   & \hspace*{0.75cm} = \;\frac{1}{2}\pi^{\frac{d}{2}}\Gamma\left(\frac{- d + \sum\nolimits^3_{i=1} \Delta_i}{2}\right) C_{\Delta_1,0}C_{\Delta_2,0}C_{\Delta_3,0} \frac{\Gamma\left(\frac{\Delta_1+\Delta_2-\Delta_3}{2}\right)\Gamma\left(\frac{\Delta_1+\Delta_3-\Delta_2}{2}\right)\Gamma\left(\frac{\Delta_2+\Delta_3-\Delta_1}{2}\right)}{\Gamma\left(\Delta_1\right)\Gamma\left(\Delta_2\right)\Gamma\left(\Delta_3\right)}.
\end{align}

\bibliography{refs}
\bibliographystyle{JHEP}

\end{document}